\documentclass{article}
\usepackage{amsmath}
\usepackage{graphicx}
\usepackage{amssymb}
\usepackage{subfigure}
\usepackage[margin=1in]{geometry}
\usepackage{tabu}
\usepackage{cite}
\usepackage{caption}

             \newcommand{\gsim}{\lower.7ex\hbox{$\;\stackrel{\textstyle>}{\sim}\;$}}
\newcommand{\lsim}{\lower.7ex\hbox{$\;\stackrel{\textstyle<}{\sim}\;$}}
\def\beq {\begin{equation}}
\def\eeq {\end{equation}}
\def\bea {\begin{eqnarray}}
\def\eea {\end{eqnarray}}
\def \PMET{\rm p{\!\!\!/}_T}

\def \PMETV{\overrightarrow{\rm p}{\!\!\!\!/}_T}
\def \PTV{\overrightarrow{\rm p}{\!}_T}
\newcommand{\br}{\begin{eqnarray}}
\newcommand{\er}{\end{eqnarray}}
\newcommand{\be}{\begin{equation}}
\newcommand{\ee}{\end{equation}}
\newcommand{\bec}{\begin{center}}
\newcommand{\eec}{\end{center}}

\usepackage{color}
\usepackage{hyperref} 
\hypersetup{linktocpage=true}
\usepackage[all]{hypcap}
\hypersetup{
   bookmarks=true,         
   unicode=false,          
   pdftoolbar=true,        
   pdfmenubar=true,        
   pdffitwindow=false,     
   pdfstartview={FitH},    
   pdftitle={My title},    
   pdfauthor={Author},     
   pdfsubject={Subject},   
   pdfcreator={Creator},   
   pdfproducer={Producer}, 
   pdfkeywords={keyword1} {key2} {key3}, 
   pdfnewwindow=true,      
   colorlinks=true,       
   linkcolor=blue,          
   citecolor=magenta,        
   filecolor=magenta,      
   urlcolor=cyan           
}

\begin{document}
\begin{flushright}
\tt{TIFR/TH/15-08},~\tt{LPSC-15/081}
\end{flushright}

\thispagestyle{empty}
\vspace*{-22mm}
\vspace*{10mm}
\hypersetup{backref=true,bookmarks}

\vspace*{10mm}

\begin{center}
{\Large {\bf\boldmath 
Probing the NMSSM via Higgs boson signatures from stop cascade decays at the LHC}}

\vspace*{10mm}

{\bf Amit Chakraborty$\rm ^a$, Dilip Kumar Ghosh$\rm ^b$, 
Subhadeep Mondal$\rm ^c$, \\
 Sujoy Poddar$\rm^d$, Dipan Sengupta$\rm ^e$}\\
\vspace{5mm}

{
$\rm ^a$ Department of Theoretical Physics, Tata Institute of 
Fundamental Research,\\
1, Homi Bhabha Road, Mumbai-400\,005, India.
\vspace{2mm}

$\rm ^b$ Department of Theoretical Physics, Indian Association for the 
Cultivation of Science, \\ 
2A \& 2B, Raja S.C.\,Mullick Road, Jadavpur, Kolkata 700\,032, India. 
\vspace{2mm}

$\rm ^c$ Harish-Chandra Research Institute, 
Chhatnag Road, Jhusi, Allahabad 211019, India.
\vspace{2mm}

$\rm^d$Netaji Nagar Day College, 
170/436 N.S.C. Bose Road, Kolkata 700092, India.
\vspace{2mm}

$\rm^e$ Laboratoire de Physique Subatomique et de Cosmologie, 
Universite Grenoble-Alpes, CNRS/IN2P3,\\
 53 Avenue des Martyrs,
F-38026 Grenoble Cedex, France.
}

\vspace*{10mm}

{\bf Abstract}\vspace*{-1.5mm}\\
\end{center}

Higgs signatures from the cascade decays of light stops are an interesting
possibility in the next to minimal supersymmetric standard model
(NMSSM). We investigate the potential reach of the light 
stop mass at the 13 TeV run of the
LHC by means of five NMSSM benchmark points where this signature is
dominant. These benchmark points are compatible with current Higgs
coupling measurements, LHC constraints, dark matter relic density and
direct detection constraints. We consider single and di-lepton search
strategies, as well as the jet-substructure technique 
to reconstruct the Higgs bosons. We find that one can probe stop masses up to 1.2 TeV
with 300 $\rm fb^{-1}$ luminosity via the di-lepton channel, while with the
jet-substructure method, stop masses up to 1 TeV can be probed with
300 $\rm fb^{-1}$ luminosity. We also investigate the possibility of the
appearance of multiple Higgs peaks over the background in the fat-jet
mass distribution, and conclude that such a possibility is viable only
at the high luminosity run of 13 TeV LHC.

%
%
\section{Introduction}
\label{sec:intro}
The discovery of a Higgs-like particle with mass close to $125~{\rm GeV}$ by both 
the ATLAS \cite{Aad:2012tfa} and CMS \cite{Chatrchyan:2012ufa} collaborations 
has led us to a new crossroad. Measurements of its spin, CP quantum number and 
couplings to the standard model (SM) particles so far are 
consistent with SM predictions for the Higgs boson
\cite{Aad:2014aba,Khachatryan:2014jba}. The natural question is whether 
this Higgs-like particle is also one of the CP-even Higgs bosons 
of the supersymmetric (SUSY) models (for a review of supersymmetry, 
see Refs.\cite{Nilles:1983ge,Haber:1984rc,Martin:1997ns}). 
The primary goal of the run-II of the LHC therefore is 
to look for new particles as well as  investigate the properties and 
couplings of this Higgs boson more precisely.

In the 
 minimal supersymmetric
standard model (MSSM), in order to push the 
light CP-even Higgs boson mass up to $125~{\rm GeV}$, a 
sizeable amount of quantum correction is required. The 
dominant contribution comes from the light third generation 
squarks, namely the stops. However, to achieve a 
$125~{\rm GeV}$ Higgs boson in the MSSM one requires to have maximally 
mixed or heavy stops (close to 1 TeV). This requirement, 
 along with the lower limits imposed on the different SUSY 
particles (specially the gluino and the light stop) after the 8 {\rm TeV} 
run at the LHC have pushed 
the fine-tuning problem of the SM to an uncomfortable level \cite{Papucci:2011wy}. 
The next to minimal 
supersymmetric standard model (NMSSM) \cite{Ellwanger:2009dp} 
provides a solution to this problem. 
The model extends the 
MSSM field content by the addition of a gauge singlet 
chiral superfield ($\rm \widehat S$). As a consequence of 
the coupling of this new gauge singlet superfield $\rm \widehat S$ 
with the Higgs doublet superfields $\rm \widehat H_u$ and $\rm \widehat H_d$,
the $\rm \lambda \widehat H_{u} \widehat H_{d} \widehat S$ term in the NMSSM 
superpotential, the tree level Higgs boson mass receives an extra tree level 
contribution which is proportional to the square of the singlet-doublet 
coupling $\rm \lambda$. Therefore a relatively smaller radiative correction 
is required in the NMSSM to obtain a $125~{\rm GeV}$ 
Higgs. This ensures that the fine-tuning issue is diluted to some
extent in the NMSSM compared to the MSSM \cite{BasteroGil:2000bw,King:2012is,Dermisek:2005ar,Ellwanger:2011mu,Ross:2011xv,Ross:2012nr}.

The phenomenology of the NMSSM is much richer than MSSM due to the presence of the additional superfield $\widehat S$. 
The Higgs sector of the NMSSM consists of five neutral 
(three CP-even $H_{1,2,3}$ and two CP-odd $A_{1,2}$) and two 
charged  ($H^{\pm}$) scalars.  
The neutral component of the scalar part of the gauge singlet 
superfield $\widehat S$ mixes with the neutral components of the 
Higgs fields, and thus there can be a significant doublet-singlet mixing after the Higgs  
mass matrix is diagonalized. For certain choices of the model parameters, one 
can find a light Higgs boson which is predominantly singlet like.
 Therefore regardless of the CP properties,
its coupling with the SM gauge bosons will be highly suppressed. 
Therefore the constraints imposed by LEP cannot exclude such a singlet like light Higgs 
boson. Thus having additional Higgses in 
the vicinity of the SM-like 125 ${\rm GeV}$ Higgs boson
is an attractive possibility in the NMSSM \cite{Belanger:2012tt,Ellwanger:2013ova,Barbieri:2013nka,Cerdeno:2013cz,
Kang:2013rj,Badziak:2013bda}. 
 While the spin, parity and coupling measurements 
of the discovered Higgs boson indicate that it is the SM-like Higgs,
 it could very well be the second lightest CP-even Higgs 
boson of the NMSSM, with SM like couplings to fermions and gauge bosons.

The Higgs bosons in the NMSSM can arise from the cascade decays of stops and 
heavy neutralinos.
A wide range of papers have studied the production of Higgs bosons and
 neutralino cascade decays to Higgs bosons in the NMSSM, 
and we refer some of them in \cite{Franke:1995tf,Ellwanger:1997jj,Choi:2004zx,
Cheung:2008rh,Stal:2011cz,Das:2012rr,Cerdeno:2013qta}.
The prospect of having additional 
NMSSM Higgs bosons in the vicinity of the 125 $\rm GeV$ Higgs boson 
was considered in \cite{King:2012is}, where the authors carefully 
analyzed the interplay of NMSSM Higgs bosons and the stop sector 
and concluded that one could achieve a modest level of fine 
tuning with a light stop below 1 ${\rm TeV}$. 
Moreover, the 
possibility of the production of multiple Higgses in NMSSM was also 
considered in \cite{Ellwanger:2014hca}, where the authors 
considered pair production of SM like Higgs bosons from 
squark and gluino cascade decays.

In a R-parity conserved SUSY scenario, the lightest  neutral, stable 
SUSY particle (LSP) serves as a good cold dark matter (DM) candidate.
The presence of a singlet superfield in the NMSSM, that gives 
rise to an extra singlino component in the neutralino sector, results 
in a very interesting DM phenomenology.
Moreover unlike the MSSM, a very light DM matter candidate as favored by 
some of the DM experiments \cite{Bernabei:2008yi,Aalseth:2010vx} can 
also be  accommodated in NMSSM. Therefore a comprehensive 
phenomenological study of the interplay of the 
Higgs sector and the neutralino sector in the NMSSM is of 
prime importance at the LHC.

At this point it is pertinent to discuss some of the 
LHC analyses dedicated to stop searches at the 8 {\rm TeV} run of LHC
relevant to this study.
Most of these searches by the ATLAS and CMS collaborations 
have been conducted in the framework of simplified topologies. 
The CMS collaboration has studied the prospect of  discovering 
the light $\rm \tilde{t}_{1}$ assuming 100\% branching ratio 
(BR) for $\rm \tilde{t}_{1}\to t\chi_{1}^{0}$, and placed the limit 
on the stop mass up to 650 ${\rm GeV}$ for massless neutralinos in the 
single lepton + jets + $\PMET$ channel
 \cite{Chatrchyan:2013xna}. The decay of the heavier stop  ($\rm \tilde t_{2}$) 
to $\rm \tilde t_{1}$ and the SM-like Higgs
($\rm \tilde{t}_{2}\to \tilde{t}_{1}H_{SM}$)
has also been studied by the CMS collaboration \cite{Khachatryan:2014doa}.
The limits at 8 {\rm TeV} are rather weak, with the heavier stop up to 575 $\rm GeV$ 
and the lighter stop up to 400 $\rm GeV$ being ruled out with 19.5 $\rm fb^{-1}$ 
of integrated luminosity. Additionally, in the framework of 
gauge mediated supersymmetry breaking (GMSB), the channel
 $\rm \tilde{t}_{1}\to t\chi_{1}^{0}$, followed by
the decay of $\rm \chi_{1}^{0}$ to $\rm H_{SM}$ and a Gravitino ($\tilde{G}$) 
was studied in \cite{Chatrchyan:2013mya}. 
The limits placed on the light stop mass for this particular scenario
 is about 400 $\rm GeV$. Furthermore, the pair production of chargino and neutralino
with the decay $\rm \chi_{2}^{0}\to \chi_{1}^{0}H_{SM}$ has also been studied 
by the ATLAS collaboration, assuming degenerate masses for
 $\rm \chi_{1}^{\pm}$ and $\rm \chi_{2}^{0} $. The limit on  
$\rm m_{\chi_{2}^{0}}$  was set to 250 GeV \cite{Aad:2015jqa}. 
 It has to be remembered 
that most of these experimental studies have been performed 
assuming that the branching ratios for all the topologies are 100\%. 
A few phenomenological studies with boosted top and 
Higgs have also been performed for 14 $\rm TeV$ run of LHC 
in the simplified model scenarios and in the context of some 
specific models as well. In particular, the prospects of $\rm \chi_{1}^{\pm}\chi_{2}^{0}\to H_{SM}W\chi_{1}^{0}$ with boosted 
Higgs was studied in \cite{Ghosh:2012mc} in the context of the constrained minimal supersymmetric 
standard model (CMSSM).  It was concluded that $\rm \chi_{1}^{\pm},\chi_{2}^{0}$ masses up to
400 GeV could be probed at 14 {\rm TeV} LHC in this framework.  The decay of heavier stops to 
$\rm \tilde{t}_{2}\to\tilde{t}_{1}H_{SM}$ was considered in \cite{Ghosh:2013qga}. The author concluded 
that $\rm \tilde{t}_{2}$ masses up to 1 {\rm TeV} could be probed at 14 {\rm TeV} with 100 $\rm fb^{-1}$ luminosity, 
in a simplified model scenario.

In this paper, we consider the production of Higgs bosons from the 
cascade decay of the $\rm \tilde{t}_{1}$ in the NMSSM framework. 
We investigate the prospects of light stop pair production followed by the decay 
of $\rm \tilde{t}_{1}$ to a top quark and any one of the heavier 
neutralinos ($\rm \chi_{i=2,..,5}^{0}$), which subsequently decays to 
one of the neutral Higgs 
states ($\rm H_1$,$\rm  H_{\rm SM} \equiv H_{2}$, $\rm A_1$)\footnote{Throughout 
this paper we will denote the SM-like Higgs 
as $\rm H_{SM}\equiv H_{2} $ and the rest of the Higgs spectrum contains $\rm H^{\pm},~H_{1,3},~A_{1,2}$.} and the 
lightest neutralino, i.e., $\rm \chi_{i}^0\rightarrow\chi_1^0 + H_{1,SM}/A_{1}$. 
The produced Higgs in this case will be boosted if there is 
a significant mass gap between $m_{\chi_{i}^{0}}$ and $m_{\chi_{1}^{0}}$.
Here we consider five benchmark points (BP) compatible with the 
current Higgs coupling measurements, DM relic density constraint, 
and direct detection constraints. The main characteristic feature 
of all these BPs is that the decay of the heavier neutralinos to 
the Higgses is dominant. We apply the 
jet substructure technique to reconstruct these boosted 
Higgs bosons to estimate the reach of stop searches in the context of 
13 $\rm TeV$ run of LHC. We also analyze the possibility of 
the appearance of multiple Higgs peaks over the background in 
the fat-jet mass distribution. It is observed that the appearance of such peaks
are only viable at very high luminosity run of LHC. However it is 
worth pursuing such a possibility to reveal the presence physics beyond the SM. 
We also compare our results with the generic lepton(s), jets + missing
energy searches at the LHC.

The rest of the paper is organised as follows. In section \ref{sec2} we discuss the
choice of benchmark points, the relevant constraints imposed and briefly review 
some of the literature relevant to this study. 
In section \ref{sec3}, we discuss our collider analysis, while in 
section \ref{sec4} we present our results and finally we conclude 
in section \ref{sec5}.    

\section{Benchmark points and constraints}
\label{sec2}  

As  already discussed, we focus on the production of the 
Higgs bosons from the cascade decay of the $\rm \tilde{t}_{1}$.
Thus we start with $\rm \tilde{t}_{1}$ pair 
production followed by the cascade decay of the stop to 
neutralinos and Higgs.  The final state therefore contains 
 top quarks, Higgs bosons, and LSPs. 
The  Higgs  decays dominantly to pair of b-quarks, while we let the
top quark decay inclusively. We present a sample Feynman 
diagram of the process of our interest in Fig.~\ref{process}. 
Few representative cascade decay modes are shown below, where 
$X$ collectively denotes the decay products of the top quarks and also 
other final state particles except the Higgses:
%

$$
\begin{array}{ccccccccc}
\rm p  p & \to &  \rm \widetilde{t}_{1} \, \widetilde{t}_{1}^{*} & \to 
&  \rm t \, \bar{t} \, \chi_{2}^{0} \, 
\chi_{2}^{0} & \to &  \rm t \, \bar{t}+ 2 H_{SM} +
2 \chi_{1}^{0}   
& \to 
&  (\rm  b \, \bar{b}) (\rm  b \, \bar{b}) + \PMET + X  \\
\rm p  p & \to &  \rm \widetilde{t}_{1} \, \widetilde{t}_{1}^{*} & \to 
&  \rm t \, \bar{t} \, \chi_{2}^{0} \, 
\chi_{2}^{0} & \to &  \rm t \, \bar{t}+ H_{SM} H_{1}  + 
2 \chi_{1}^{0}   
& \to 
&  (\rm  b \, \bar{b}) (\rm  b \, \bar{b}) + \PMET + X  \\
\rm p  p & \to &  \rm \widetilde{t}_{1} \, \widetilde{t}_{1}^{*} & \to 
&  \rm t \, \bar{t} \, \chi_{2}^{0} \, 
\chi_{3}^{0} & \to &  \rm t \, \bar{t}+ H_{SM} H_{1}  + 
2 \chi_{1}^{0}   
& \to 
&  (\rm  b \, \bar{b}) (\rm  b \, \bar{b}) + \PMET + X  \\
\rm p  p & \to &  \rm \widetilde{t}_{1} \, \widetilde{t}_{1}^{*} & \to 
&  \rm t \, \bar{b} \, \chi_{2}^{0} \, 
\chi_{1}^{-} & \to &  \rm t \, \bar{b}+ W^{-} H_{SM}/H_{1}/H_{3} +
2 \chi_{1}^{0}   
& \to 
& (\rm   b \, \bar{b}) + \PMET + X.  \\
\label{eq1}
\end{array}
$$

\begin{figure}[!htb]
\begin{center}
\includegraphics[angle =0, width=0.5\textwidth]{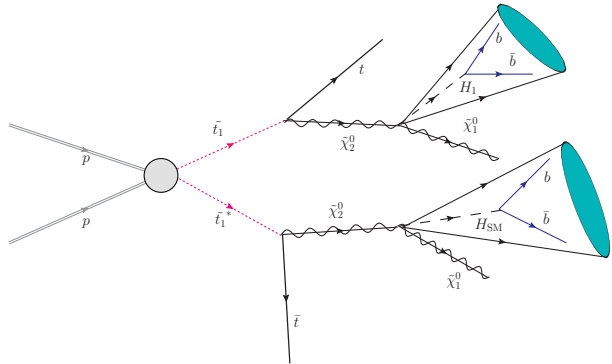}
\caption{ { A sample Feynman diagram for the channel under consideration. }}
\label{process}
\end{center}
\end{figure}


%

The decay of the next-to lightest supersymmetric 
particle (NLSP) to LSP + X ($\rm X=Z/H_{SM}$) in 
the NMSSM framework has been already addressed in 
\cite{Ellwanger:2014hia}, where the authors pointed out that 
for a singlino like LSP satisfying DM constraints, this 
decay would lead to a reduction of missing transverse 
energy making it difficult to probe at the LHC. 
The authors also looked at the prospects of Higgs production from 
the cascade decay of squarks and noted that small 
values of $\lambda$ and $\kappa$ were required to 
satisfy the invisible Higgs decay constraint by suppressing the mixing 
of the singlet like $\chi_{1}^{0}$ with Higgsinos, for cases 
where $m_{\chi_{1}^{0}} \sim$~5 GeV. The authors 
showed that the $\rm jjb\bar{b}$ background was significantly 
large at 13 {\rm TeV} LHC and so they proposed that tagging multiple 
b-jets  along with the requirement of a large number of jets can be useful 
to reduce the large background. The decay 
$\rm \chi_{2,3}^{0}\to\chi_{1}^{0}H_{SM}$ was also discussed in \cite{Dutta:2014hma}, 
where the authors also argued along the same lines. 
The authors looked at  $\rm p p\to\chi_{i}^{0}\chi_{j}^{0}jj$,
with subsequent decay of the neutalino states to the SM-like Higgs. 
They showed that four b-tagged jets in addition to $\rm \PMET$ 
can be a viable channel at 14 {\rm TeV} LHC. The authors also suggested 
ways of distinguishing MSSM from NMSSM by observing dark matter patterns 
and neutralino branching ratios at the colliders. This 
particular region of parameter space specific to the NMSSM has 
therefore received significant attention. 
From the above studies it is also evident that probing such a channel 
at the LHC environment is a very difficult task. A detailed 
collider study at  13 {\rm TeV} LHC is preformed here.

The benchmark points for this study are chosen with the intention of 
probing the regions of parameter space where the decay of 
$\rm \chi_{2,3}^{0}$ to LSP + Higgs is dominant. We present our 
five representative benchmark points in Table \ref{tab1}. 
The branching ratios of the relevant decay modes 
and some of the crucial observables corresponding to the 
benchmark points are tabulated in 
Table \ref{tab2}. The particle spectrum and decays are 
generated using NMSSMTOOLS-4.2.1\cite{Ellwanger:2005dv}. 
Since we are interested in the decay of the light stop, we decouple the rest of 
the spectrum, i.e, we set the mass of the gluino and the rest of the 
squarks and sleptons at 3 {\rm TeV}. Furthermore, we want the light stop 
to be predominantly left handed and therefore we set the 
soft masses for the right handed 3rd generation squarks to be at 
3 {\rm TeV}, so that the decays $\rm \tilde{t}_{1}\to t \chi_{2,3}^{0}$ 
and $\rm \tilde{t}_{1}\to b \chi_{1}^{\pm}$ are dominant. 
It is pertinent to note that a detailed collider analysis 
for the left handed light stop was performed by some of the 
same authors in an earlier work using the boosted top technique in 
the context of phenomenological MSSM \cite{Chakraborty:2013moa}. 
Since our primary channel of interest for the subsequent stage 
of the cascade is $\rm \chi_{2,3}^{0}\to\chi_{1}^{0}H_{SM}/H_{1}/A_{1}$, we allow top 
to decay inclusively, while Higgs decays dominantly to $b\bar b$. 
 It has 
to be remembered that the whole scenario is additionally 
complicated by the fact that one has to satisfy the DM 
relic density and also DM direct detection constraints, 
if $\chi_{1}^{0}$ is assumed to be a valid DM candidate.

In order to ensure that $\rm \chi_{2,3}^{0}$
has a large higgsino component, we fix 
$\rm M_1$ and $\rm M_2$ to 1.5 {\rm TeV} for all benchmark 
points and vary 
$\rm \mu_{eff}$ in the range  of 260-280 {\rm GeV}.
This also ensures that $\rm \chi_{4,5}^{0}$ are predominantly 
gaugino like and have mass close to 1.5 TeV. Thus they play 
no role in the cascade decay of the light stop.
 On the other hand, 
the large value of $\rm \lambda$ ensures that 
the lightest neutralino is predominantly singlino for 
all of the parameter space points. The two 
decay modes that compete here are 
$\rm \chi_{2,3}^{0}\to\chi_{1}^{0} Z$ and 
$\rm \chi_{2,3}^{0}\to\chi_{1}^{0} H_{1,SM}$, when the 
mass difference, $\rm m_{\chi_{2,3}^{0}}-m_{\chi_{1}^{0}} > m_{H_{1,SM}}$ 
is satisfied. In the MSSM, this competition is predominantly 
determined by the gaugino-higgsino components of the neutralino, 
such that the decay of heavier neutralino states to lighter 
ones and Higgs has full gauge strength if one of the neutralino is 
gaugino like while the other is higgsino like. However, in the 
NMSSM, the LSP can have a significant amount of singlino admixture, 
and this is the case for our all the benchmark points. It is 
known that the Z boson only couples to the higgsino like component 
of the lightest neutralino and therefore the 
$\rm \chi_{1}^{0}\chi_{2}^{0}Z$ coupling is suppressed for a 
dominantly singlino like neutralino. Hence 
$\rm \chi_{2}^{0}\to \chi_{1}^{0}H_{SM}$ dominates in all our benchmark 
points (see Table \ref{tab2}).
Here, we perform a 
small scan over the NMSSM parameter space 
and consider the parameters presented in Table \ref{tab1} such 
that the relevant decay modes are dominant and all the present low energy 
constraints are satisfied. In the Higgs sector, we ensure that 
the LEP constraints as well as SM Higgs coupling constraints at the 
end of 8 {\rm TeV} run of the LHC are satisfied \cite{Beringer:1900zz,Aad:2014aba,Khachatryan:2014jba}. 
The lightest Higgs boson being dominantly 
singlet like, the $\rm H_{1} ZZ$ coupling is suppressed 
ensuring that the LEP bounds are easily satisfied. 
A consequence of the singlet like $\rm H_1$ is that the 
branching ratio of $\rm \chi_{2,3}^{0}\to \chi_{1}^{0}H_{1}$ is 
rather small (see Table \ref{tab2}). We also note 
that $\rm M_{A}$, the pseudoscalar mass parameter, plays 
an important role to satisfy the DM constraints. For the parameter 
space of our interest, the LSP annihilates dominantly via the pseudoscalar, 
and therefore a small tuning in the mass of $\rm M_{A}$ is required to obtain the 
DM relic density within 2$\sigma$ of WMAP/PLANCK data 
\cite{Hinshaw:2012aka,Ade:2013zuv}. Additionally, we make 
sure that the benchmark points of our choice satisfies the direct 
detection cross-section bounds obtained from the 
LUX \cite{Akerib:2013tjd} collaboration.

\begin{table}[ht!]
\small
\begin{center}
\tabulinesep=1.2mm
\begin{tabu}{|c|c|c|c|c|c|}
\hline
  &  P1 & P2 & P3 & P4 & P5  \\
\hline
$\rm m_{\widetilde{Q_3}}$   & 830  & 920 & 1000  & 1180 & 1350   \\
\hline
$\rm \tan\beta$   & 10.2  & 12.2  & 10.4  & 10.3 & 10.2   \\
\hline
$\rm \lambda$   & 0.34  & 0.15 &  0.52 & 0.6 & 0.44  \\
\hline
$\rm \kappa$   &  0.06 &   0.03  & 0.09 & 0.09 & 0.07  \\
\hline
$\rm A_{\lambda}$   & 2700  & 2700 & 2700  &2700  &  2700   \\
\hline
$\rm A_{\kappa}$   &  -1200 & -1200  & -1200 & -1200 &-1200   \\
\hline
$\mu_{eff}$ &  280 & 264 &  262 & 268 & 277   \\
\hline
$m_{H_{SM}}$ &124.3 & 124.9 & 124.3& 125.1 & 124  \\
\hline
$m_{H_1}$ & 67.1& 67.1&62.3 & 73.8 &67.2 \\
\hline
$m_{A_1}$ & 214 & 226.8 & 194.4& 154.0 &193 \\
\hline
 $\rm m_{\widetilde{t}_{1}}$  & 804.2 & 908.2 & 1003.7  & 1211  & 1392.5  \\
\hline
$\rm m_{\widetilde{b}_{1}}$  & 821.4  &923.1  & 1018 & 1226 & 1408.6   \\
\hline
$\rm m_{{\chi}_{1}^{0}}$  & 98.6 & 107.2 & 87.3 &  77.6 & 87.4  \\
\hline
$\rm m_{{\chi}_{2}^{0}}$  & 290.1 & 268.0 & 282.7 & 293.2 & 292.4  \\
\hline
$\rm m_{{\chi}_{3}^{0}}$  & 293.5  & 273 & 283.1 & 294.2 & 295.1  \\
\hline
\hline
$\rm m_{{\chi}_{1}^{\pm}}$  & 284.7 & 268.5 & 266.5 & 273.2 & 282.5   \\
\hline
\end{tabu} 
\caption{ The parameters and masses for the five benchmark points. 
All the other parameters are set to their fixed values as described in the text. 
All masses are in {\rm GeV} unit.}
\label{tab1}
\end{center}
\end{table}

\begin{table}[ht!]
\small
\begin{center}
\tabulinesep=1.2mm
\begin{tabu}{|c|c|c|c|c|c|}
\hline
  &  P1 & P2 & P3 & P4 & P5   \\
\hline
$\mu^{\gamma\gamma}_{h_2} (ggF)$, $\mu^{\gamma\gamma}_{h_2} (VBF/VH)$   & 0.95, 0.96  &0.94, 0.94  & 0.96, 0.96  & 0.92, 0.91 & 0.92, 0.91 \\
\hline
  $\mu^{\tau\tau}_{h_2} (ggF)$, $\mu^{\tau\tau}_{h_2} (VBF/VH)$  & 1.0, 0.99  & 0.99, 0.97 & 0.99, 0.98  &1.0, 1.0  & 1.0, 1.0  \\
\hline
 $\mu^{bb}_{h_2} (ttH)$, $\mu^{bb}_{h_2} (VBF/VH)$   & 1.0, 0.97  & 0.95, 0.95 & 1.0, 1.0  &1.01, 0.93  & 1.0, 1.0 \\
\hline
 $\Omega h^2$, $\sigma_{SI}$($\times 10^{47},~{\rm cm^{2}}) $ & 0.123, 6.4       &0.122, 1.9   & 0.115, 8.1   & 0.125, 1.8  & 0.126, 9.3  \\
\hline
$\rm BR(\widetilde{t}_{1} \to t \, {\chi}_{2,3,4}^{0}) (\%)$ & 92  & 94 & 97  & 81 & 88  \\
\hline
$\rm BR(\widetilde{t}_{1} \to b \, {\chi}_{1,2}^{\pm})(\%)$ & 7 & 6 &3 &4 &4 \\
\hline
$\rm BR({\chi}_{2}^{0} \to {\chi}_{1}^{0} \rm{H_{1,SM}})(\%)$  &6, 60 &8, 61 &20, 57 &18, 49 &14, 54 \\
\hline
$\rm BR({\chi}_{3}^{0} \to {\chi}_{1}^{0} \rm{H_{1,SM}})(\%)$  & 3, 14 &8, 1 & 3, 15&6, 16 &4, 16.3 \\
\hline
$\rm BR(H_{1}\to b\bar{b})(\%)$ & 86&69 &91 &90 & 85 \\
\hline
$\rm BR(H_{2}\to b\bar{b})(\%)$ &64 &64 &64 &64 &66 \\
\hline
$\rm BR(A_{1}\to b\bar{b})(\%)$ &1 &1 & 1& 90& 1.5 \\
\hline
\end{tabu} 
\caption{ Some of the Higgs signal strengths, dark matter relic 
density, direct detection constraints  and the relevant branching ratios 
for the particular study.}
\label{tab2}
\end{center}
\end{table}

All the benchmark points satisfy the following set of constraints:

\begin{itemize}

\item The mass of the SM-like Higgs must be within the range $\rm 123~ {\rm GeV} <M_{H_{SM}} < 129~ {\rm GeV}$. 
Due to the small difference in the central values of the ATLAS and CMS 
measurements and also considering the theoretical uncertainty 
in Higgs mass calculation, we use 126 $\pm$ 3 {\rm GeV} as a conservative 
estimate \cite{Degrassi:2002fi,Goodsell:2014pla}.
  
\item The SM-like Higgs couplings to be within 2$\rm \sigma$ of the measured 
values provided by the ATLAS and CMS collaborations at the end of 7 + 8 {\rm TeV} 
run of LHC with approx. 25 $\rm fb^{-1}$ of data 
\cite{Aad:2014aba,Khachatryan:2014jba}( See Table \ref{tab:higgs} for 
more details.) 

\begin{table}[!ht]
\small
\begin{center}
\tabulinesep=1.2mm
\begin{tabu}{|c|c|c|}
\hline
\multicolumn{1}{|}{} &
\multicolumn{2}{|c|}{Signal Strength ($\mu$)}  \\
\cline{2-3}
Channel  & ${\rm CMS}$ & ${\rm ATLAS}$ \\
\hline
${\gamma\gamma}$ & $1.12\pm0.24$ \cite{Khachatryan:2014jba}& $1.17\pm 0.27$ \cite{Aad:2014eha}\\
${\rm WW}$& $0.83\pm 0.21$\cite{Khachatryan:2014jba}& $1.09^{+0.23}_{-0.21}$ \cite{ATLAS:2014aga}\\
${\rm ZZ}$ & $1.0\pm 0.29$ \cite{Khachatryan:2014jba}& $1.44^{+0.40}_{-0.33}$ \cite{Aad:2014eva} \\
${\rm bb}$ & $0.84\pm 0.44$ \cite{Khachatryan:2014jba}& $0.52\pm{0.32}\pm{0.24}$ \cite{Aad:2014xzb} \\
${\tau\tau}$ & $0.91\pm 0.28$ \cite{Khachatryan:2014jba}& $1.43^{+0.43}_{-0.37}$ \cite{Aad:2015vsa}\\
\hline
\end{tabu} 
\caption{Updated results on Higgs coupling measurements by the ATLAS and 
CMS collaborations at the end of 7 + 8 {\rm TeV} run of the LHC with approx. 25 $\rm fb^{-1}$ of data. }
\label{tab:higgs}
\end{center}
\end{table}

\item The BR of the rare b-decays $\rm B \to X_{s}\gamma$ and 
$\rm B_{s} \to \mu^+ \mu^-$ are within 2$\sigma$ of the experimental 
results \cite{Amhis:2014hma}. We assume: \\
\bea
 \rm 2.77\times 10 ^{-4} < BR(B \to X_{s}\gamma) < 4.09\times 10^{-4} \nonumber \\ 
 \rm 1.0\times 10 ^{-9} < BR(B_{s} \to \mu^+ \mu^-) < 5.2\times10^{-9} \nonumber 
\eea

\item The LEP  lower bound on the chargino mass, 
${M_{\chi^{\pm}_1}} > {\rm 103.5~{\rm GeV}}$ \cite{Beringer:1900zz}.

\item We apply the upper bounds on the DM direct detection cross-section 
from LUX \cite{Akerib:2013tjd} and allow the DM relic density to be  
within 2$\sigma$ of the WMAP/PLANCK \cite{Hinshaw:2012aka,Ade:2013zuv}
data. Note that, precise value of DM relic density do not have 
any significant impact on our final results. We assume, 
\bea
0.115 \le {\rm {\Omega_{DM}h^{2}}} \le 0.126 \nonumber. 
\eea
\end{itemize}


Before we end this section, we would like to remind 
the readers that for all our benchmark points, the 
branching ratios for the decay $\rm \tilde{t}_{1}\to t\chi_{2,3}^{0}$ 
are significantly large owing to the large higgsino component 
in $\rm \chi_{2,3}^{0}$ (see Table \ref{tab2}). The dominantly 
Higgsino-like $\chi_{2}^{0}$ and largely singlino-like $\chi_{1}^{0}$ 
ensure that $\rm \chi_{2}^{0}\to\chi_{1}^{0}H_{SM}$ is 
significantly large, which is necessary for our collider strategy. 
Since $\rm H_{SM}$ is SM-like, it has a large branching ratio to $\rm b\bar{b}$, 
while $\rm H_1$, due to the large singlet component and reduced coupling to ZZ also dominantly 
decays to $\rm b\bar{b}$. From Table \ref{tab2}, it is evident that all the 
reduced signal strengths of the SM-like Higgs are well within 2$\sigma$ of the 
limits imposed from the Higgs coupling measurements at the LHC. With these comments 
on the benchmark points, we now proceed to discuss our signal and background 
collider analysis in the next section. 
 
\begin{figure}[!htb]
\begin{center}
\includegraphics[angle =0, width=0.45\textwidth]{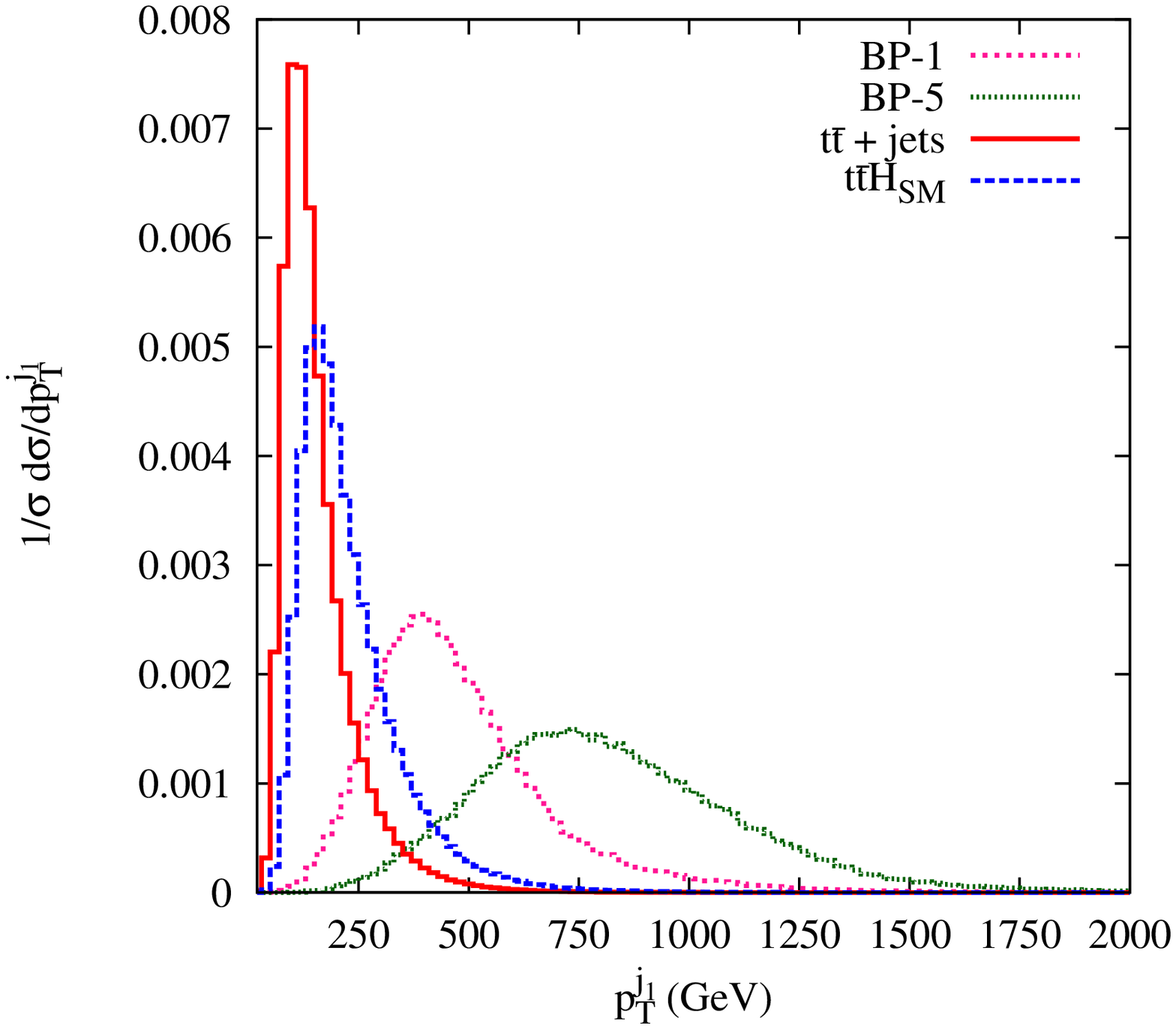}  
\includegraphics[angle =0, width=0.45\textwidth]{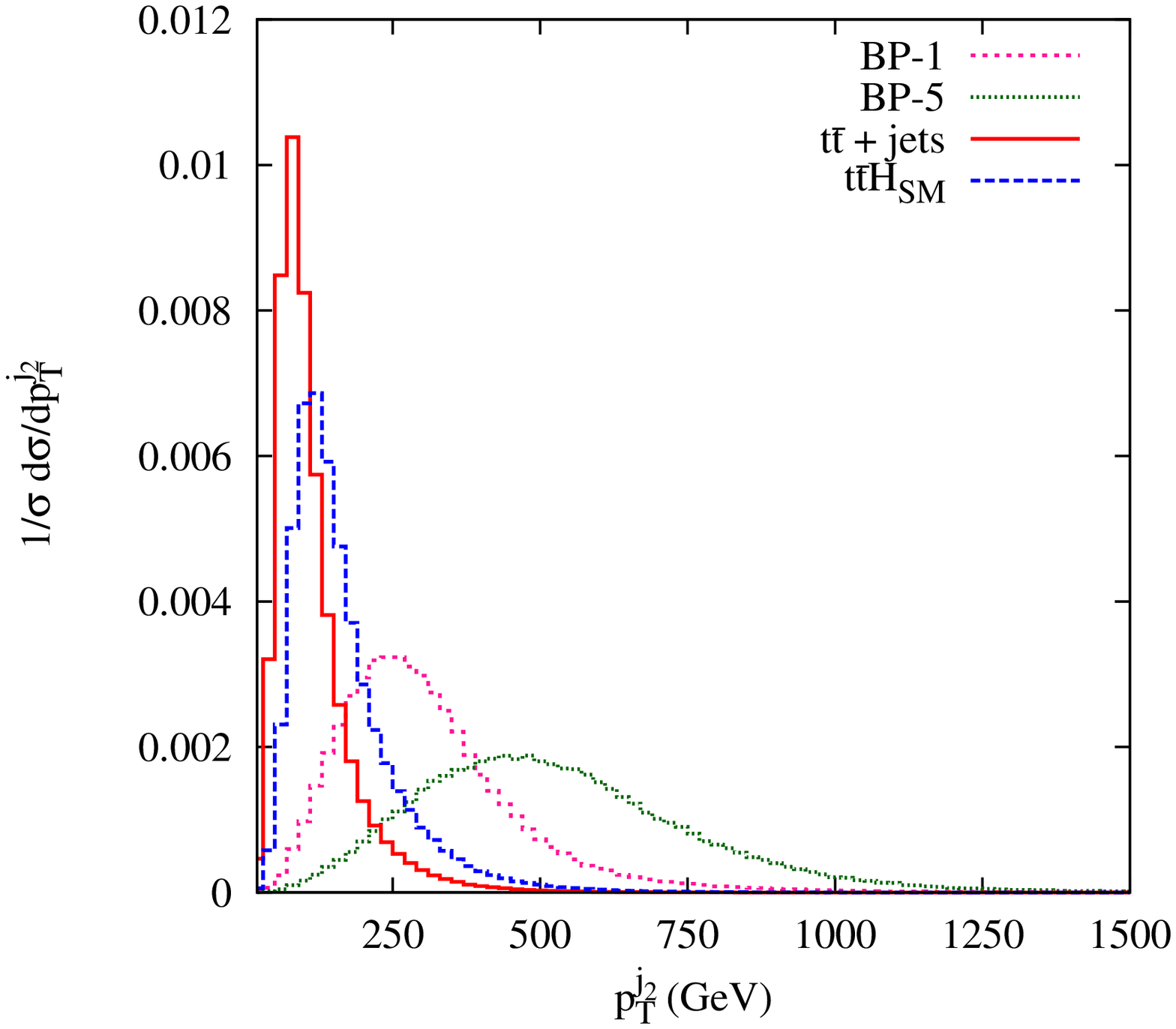} 
\vspace*{-1.5cm}
\caption{ {  $\rm p_T$-distribution of the leading two fat-jets for 13 {\rm TeV} the LHC. We reconstruct the jets 
using the C/A algorithm with jet radius R = 1.2. The benchmark points BP1 and BP5 are as 
tabulated in Table \ref{tab2}.}}
\label{ptj12}
\end{center}
\end{figure}

\section{Analysis  of the signal and background}
\label{sec3}

In this section we describe the strategy for the collider study. The entire 
analysis is divided into two parts: first we use the jet substructure 
technique to analyze the boosted Higgs scenario. This is followed by 
 the standard leptonic search strategies at the LHC. We 
then compare these two methods and discuss the merits and demerits of each of 
these two procedures.  

\subsection{Jet substructure analysis}

The Higgs tagging technique, introduced in the work of 
Butterworth-Davison-Rubin-Salam (BDRS) \cite{Butterworth:2008iy}, has 
emerged as a powerful tool to study physics in the boosted regime. 
The efficacy of the BDRS tagger 
lies in the sharp resolution of the Higgs peak in the $\rm H\to b\bar{b}$ channel. 
In the analysis, 
we apply the BDRS Higgs tagger to tag the Higgs originating from 
the decay of the heavier neutralinos. From  Table \ref{tab1}, it is 
clear that the mass difference $\rm m_{\chi_{2}^{0}}-m_{\chi_{1}^{0}}$ 
is not significantly big enough to generate a large boost to the Higgs. However, 
being part of a cascade decay, the boost to the Higgs boson  
originate from the two step decay chain of the stop. To demonstrate 
the larger boost in the signal process as compared to the background,
 we present, in Fig.~\ref{ptj12}, the $\rm p_{T}$ distribution of the leading two ``fat-jets" 
reconstructed using the Cambridge-Aachen (C/A)\cite{Dokshitzer:1997in} 
jet algorithm with the jet radius R=1.2, for 
two benchmark points BP1 and BP5, as well as the SM $\rm t\bar t + {\rm jets}$ background. 
One can observe that for the signal, the peak of the distribution is over 400 {\rm GeV}, 
while the background falls sharply from 200 {\rm GeV}. We notice that with the increase 
in the stop mass (BP1 to BP5), the fat-jet distribution moves to the higher $\rm p_{T}$
region, as expected. To optimize the Higgs tagging efficiency 
we considered the values of the fat-jet radius R to be 1, 1.2 and 1.5 and
concluded that R=1.2 is the best choice in terms of signal to background ratio. Furthermore we 
varied the jet $\rm p_{T}$ threshold, considering three 
different values 200 {\rm GeV}, 400 {\rm GeV}, 600 {\rm GeV} and arrived at the conclusion that 
a jet radius R=1.2 and a jet $\rm p_{T}$ threshold of 200 {\rm GeV} gives the best 
efficiency in terms of enhancement in the signal and reduction 
of background. Since the boost to the Higgs boson in our case 
is not significantly large, we achieve a modest Higgs tagging 
efficiency ($\epsilon \sim $ 5-6 \% for $\rm p_T \sim $ 
200 - 300 {\rm GeV}). However this is an order of magnitude larger 
compared to the largest background in terms of cross section, 
namely $\rm t\bar{t} + {\rm jets}$ (see Table \ref{tab3} and \ref{tab4}). 
An added advantage of using 
the BDRS Higgs tagger is that it 
uses a filtering technique to discard the contamination from 
the underlying events, which in turn helps to improve the 
resolution of the Higgs peak. As a consequence of the 
superior mass resolution, it is also a powerful 
tool against the backgrounds containing multiple b-jets 
like $\rm t\bar{t} + {\rm jets}$, Z($\rm \to b\bar{b}$) + jets, $\rm t\bar t b \bar b$, 
$\rm t \bar t W$, $\rm t \bar t Z$. The SM backgrounds that we 
consider are listed below,
\begin{itemize} 
\item  $\rm t\bar{t}$ + jets : The decay of top quark yield two b-jets 
in the final state and thus the reconstructed fat-jet could 
in principle fake a Higgs signal.

\item  $\rm Z(\to b\bar{b})$ + jets: In this case although the fat-jet mass 
distribution should peak at the Z boson mass there 
might be a large tail. Owing to the large Z + jets cross section, 
this could potentially be significant in the end. 

\item $\rm t\bar{t}b\bar{b}$, $\rm t\bar{t}Z$, $\rm t\bar{t}W$ : These backgrounds 
contain at least 2 b-jets, which may be tagged as a fake SM Higgs boson.

\item  $\rm t\bar{t}H_{SM}$ : This constitutes the irreducible background to our 
signal with at least one tagged Higgs jet and $\PMET$. However the Higgs 
in this case is not expected to be as boosted as the signal.
\end{itemize}

We use NMSSMTOOLS-4.2.1 \cite{Ellwanger:2005dv,Ellwanger:2006rn,Ellwanger:2004xm,Das:2011dg} to generate the NMSSM 
mass spectrum, while signal and background events are 
simulated using MADGRAPH5 \cite{Alwall:2011uj} and passed to 
PYTHIA6 (version 6.4.26) \cite{Sjostrand:2006za}
for showering and hadronization. For the signal ($\rm \widetilde{t_1} \widetilde{t^{*}_1} + 2~{\rm  jets}$) 
and $\rm t\bar{t}$/Z/W + jets (up to 2 jets), 
we perform the matrix element parton shower (ME-PS) merging 
using the MLM prescription \cite{Hoche:2006ph} with a merging 
parameter of 30 {\rm GeV}. 
We use FastJet-3.1.0 \cite{Cacciari:2011ma} for jet reconstruction, and 
implementation of our jet substructure analysis. The rare b-decay observables, 
DM relic density and direct detection cross-sections are calculated 
using NMSSMTOOLS.

We observe that even after tagging of the Higgs jet, there is still a significant 
amount of hadronic activity in the signal process. This activity 
can be used as an effective handle against the background. Therefore the 
jet reconstruction procedure in our case is a two step process. We 
first reconstruct fat-jets using the Cambridge-Aachen (C/A) 
\cite{Dokshitzer:1997in} algorithm using a jet radius of R=1.2 
with a jet $\rm p_{T}$ threshold of 200 {\rm GeV} followed by the use of 
the BDRS \cite{Butterworth:2008iy} Higgs tagger with a default mass 
drop criteria of $\rm \mu =$ 0.67 and the filtering parameter
 $\rm R_{filt}$ = $\rm min(R_{b\bar{b}}/2, 0.3)$. Once one (or multiple) 
SM-like Higgs boson(s) are tagged, we remove those particles from the fat-jet list 
and recluster the 
remaining stable hadrons using anti-$\rm k_{T}$ \cite{Cacciari:2008gp} 
jet algorithm with a jet radius of R=0.4,  $\rm p_{T} \ge 20~ {\rm GeV}$. 
We select jets with $\rm p_{T}^{j} \ge 30~{\rm GeV}$ and 
pseudorapidity $\rm |\eta_{j}|< 3$ for further 
collider analysis. Leptons (electrons and muons) are selected 
with $\rm p_{T}^{\ell} > 10~ {\rm GeV}$ 
and $\rm |\eta_{\ell}| < 2.5$. Isolation of leptons are performed by demanding 
that the sum of the scalar $\rm p_T$ of all stable visible particles 
within a cone of radius $\rm \Delta R = 0.2$ around the lepton 
should not exceed 10\% (15\% for muons) of 
$\rm p_{T}^{e}$ ($\rm p_{T}^{\mu}$). The missing transverse energy  
is constructed 
using all the stable final state visible particles 
with all jets with $\rm p_T^j > 20$~{\rm GeV} and 
$\rm |\eta_j| < 4.5$, and all leptons with $\rm p_T^\ell > 10$~{\rm GeV} and 
$|\rm \eta_\ell| < 2.5$. Following the ATLAS 
collaborations \cite{ATLAS-CONF-2014-046}, we consider a $\rm p_T$-dependent b-tagging 
efficiency  :

$$
\eta_b = 
\left\{ \begin{array}{ll} 0~~ & {\rm for} \ \ {\rm p_T^b} \leq 20~{\rm GeV} \\
                          0.6 & {\rm for} \ \ 20~{\rm GeV} < {\rm p_T^b} < 50~{\rm GeV} \\
                          0.75 & {\rm for} \ \ 50~{\rm GeV} < {\rm p_T^b} < 400~{\rm GeV} \\
                          0.5 & {\rm for} \ \ {\rm p_T} \geq 400~{\rm GeV}
               \end{array} \right.   
$$

\begin{figure}[!htb]
\begin{center}
\includegraphics[angle =0, width=0.45\textwidth]{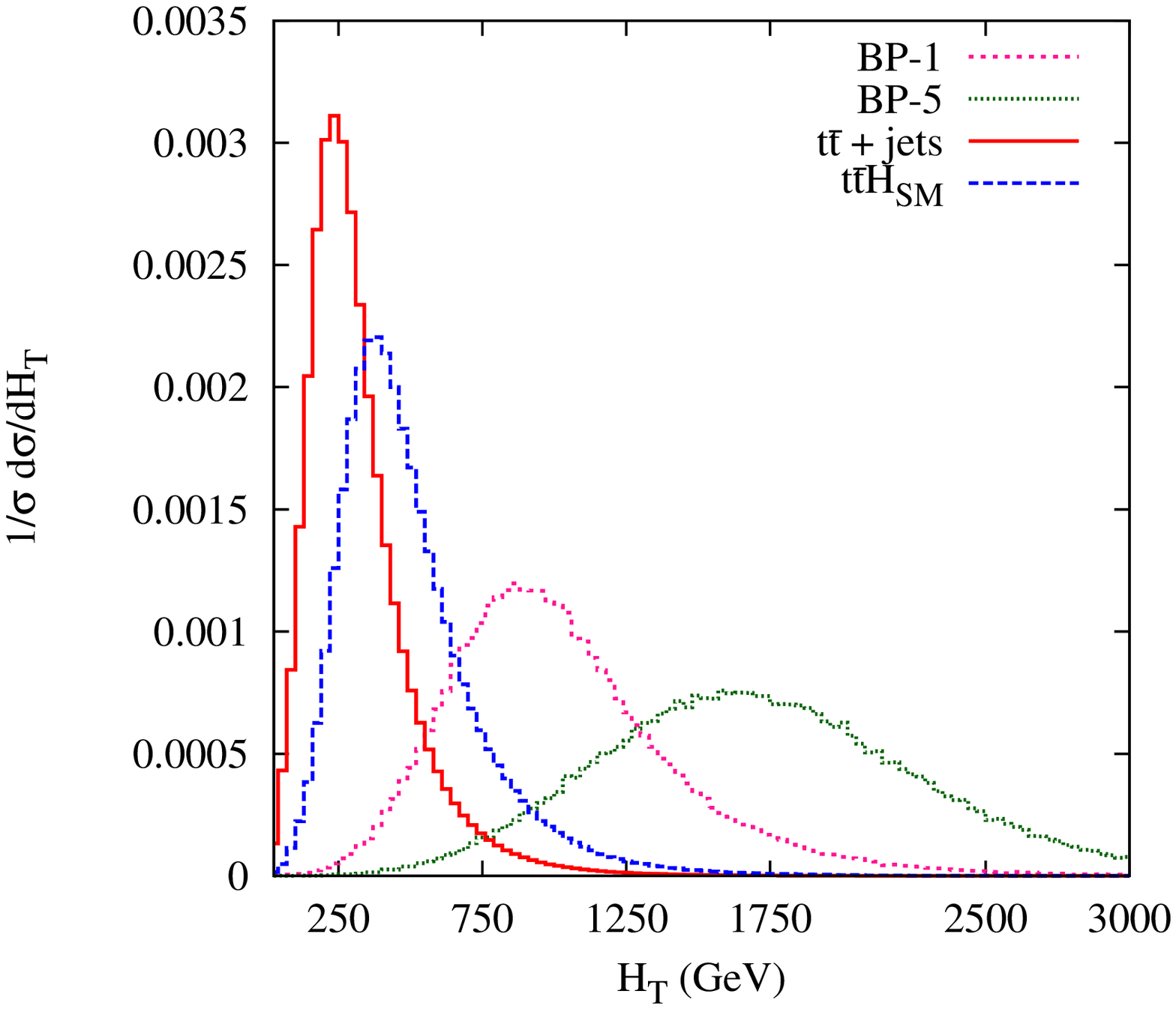} 
\includegraphics[angle =0, width=0.45\textwidth]{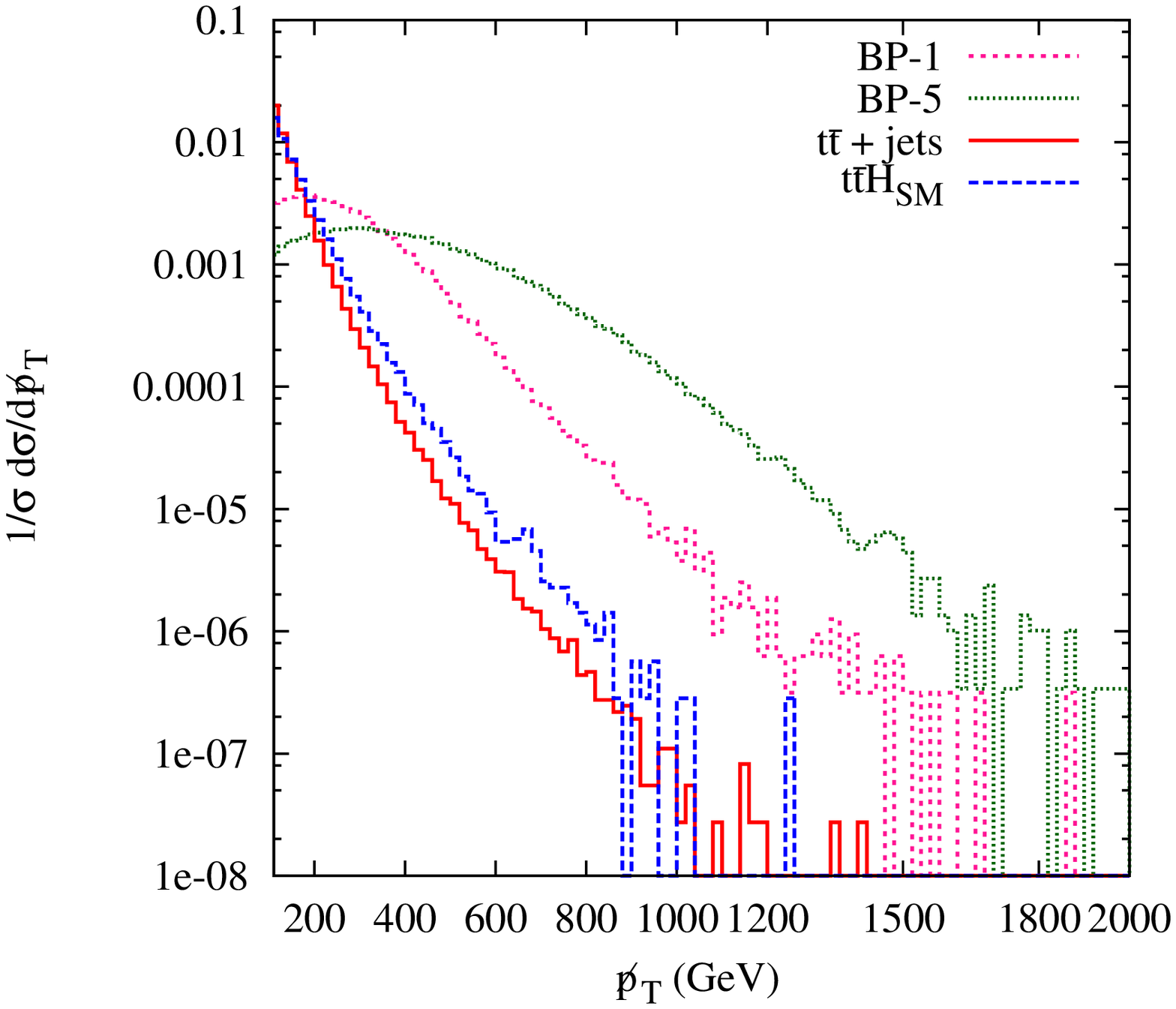}  
\vspace*{-1.5cm}
\caption{ {  Left panel shows the distribution of $\rm H_T$ where 
$\rm H_{T}=\Sigma_{j} p_{T}^{j}$ with the anti-$\rm k_T$ jets selected 
after imposing the basic Higgs tagging criteria, while 
right panel displays the distribution of the $\rm \PMET$ at the 
13 {\rm TeV} LHC. }}
\label{ptmiss_ht}
\end{center}
\end{figure}

After generating events, the following set of 
kinematic cuts are applied,

\begin{itemize}

\item C1:  {We demand that at least one of the fat-jets is identified as 
the SM-like Higgs boson with the criteria 
$\rm 110~{\rm GeV} < m_{J} < 140~{\rm GeV}$}, where $\rm {m_J}$ 
denotes the fat-jet mass. The fat-jet is deemed to be a Higgs-like jet if and only if two b-jets 
are identified inside the fat-jet. In the signal process, however, 
there are additional Higgs bosons which can be tagged and 
therefore at the end of all selection cuts we look at the 
$\rm m_{J}$ distribution, to find if additional peaks are 
visible over the background. The Higgs 
boson tagging ensures that most of the backgrounds with multiple 
b-jets in the final state are significantly reduced.   

\item C2:   {In this work we consider a hadronic final state, and therefore we impose a 
lepton veto.}

\item C3:   {For the signal, even after at least one SM-like 
Higgs boson is tagged, there are significant number of b-jets 
originating from the top quarks. However, these extra b-jets are 
unlikely to be present in the backgrounds like Z ($\rm \to b\bar{b} $) + 
jets, $\rm t\bar{t}$  + jets after satisfying the Higgs mass 
criteria. Therefore, we demand at least one additional 
b-jet in the event.}

\item C4: {An additional selection criteria of 
$\rm H_{T}=\Sigma_{j} p_{T}^{j} > 800~ {\rm GeV}$ is imposed. 
The jets in this case are the $\rm anti-k_{T}$ jets constructed 
by reclustering the rest of the hadronic sample after  removing the hadrons 
contributing to the Higgs tagged jet as mentioned earlier. The larger 
hadronic activity in the signal implies a larger value of 
$\rm H_T$ compared to the background. This cut is particularly 
useful in suppressing the irreducible  $\rm t \bar{t}H_{SM}$ background.
In the left panel of Fig.~\ref{ptmiss_ht}, we present the $\rm H_T$ 
distribution corresponding to two signal benchmark points 
BP1 and BP5, along with the two dominant backgrounds $\rm t\bar{t}$ + jets and 
 $\rm t \bar{t}H_{SM}$.}

\item C5:  {As a final selection criteria we impose a  $\rm \PMET$ 
cut of 175 {\rm GeV}. The signal $\rm \PMET$ distribution is harder 
compared to that of the SM backgrounds because of the 
heavy LSP in the final state of the signal events. In 
the right panel of Fig.~\ref{ptmiss_ht}, 
the $\rm \PMET$ distribution for the signal and $\rm t\bar{t}$ + jets 
are presented to illustrate this feature. }

\end{itemize}

\begin{figure}[!htb]
\begin{center}
\includegraphics[angle =0, width=0.45\textwidth]{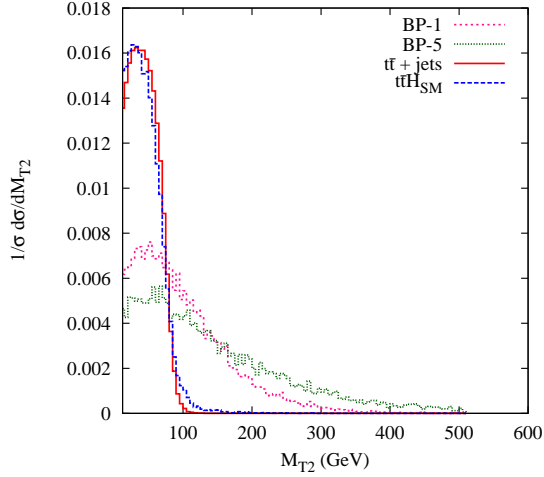}  
\vspace*{-1.5cm}
\caption{ {  The $\rm M_{T2}$ distribution at 13 {\rm TeV} LHC energy  for the benchmark points P1, 
P5 and the SM backgrounds $\rm t\bar{t}H_{SM}$ and $\rm t\bar{t}$ + jets.}}
\label{mt2}
\end{center}
\end{figure}


\subsection{Alternative analysis: Final state with leptons }

In order to understand whether the jet substructure technique is more effective 
than other search strategies, we perform a separate collider analysis 
with leptonic final states. Here we probe two such possible scenarios, 
one dedicated to the single lepton final state (denoted as Case-I), while 
other being the di-leptonic channel (denoted as Case-II).
The SM backgrounds for both these channels are 
$\rm t\bar{t}$ +jets, $\rm W(\to l\nu)$ +jets, 
Z +jets, $\rm t\bar{t}Z(\to\ell^{+}\ell^{-})$, $\rm t\bar{t}W$ (where the 
leptons can originate either from $\rm W$ or from the top) and 
the irreducible $\rm t\bar{t}H_{SM}$. To analyze these two channels, 
we devise two sets of search strategies and optimize our selection cuts 
to obtain a good signal significance at the LHC.  {We use FASTJET to 
reconstruct the jets with anti-$\rm k_{T}$ \cite{Cacciari:2008gp}
jet algorithm with a jet radius of R=0.4 and $\rm p^{\rm min}_{T} \ge 20~ {\rm GeV}$.
We select jets with $\rm p_{T}^{j} \ge 30 ~{\rm GeV}$ and $\rm |\eta_{j}|\le 3$ for further 
analysis}. Below, we list the optimized selection cuts for both the cases separately. 

\begin{enumerate}
\item {\bf{Case I (Single lepton)}} : 
\begin{itemize}
\item C6: We demand at least 4 jets with 
$\rm p_{T}^{j1} > 200~{\rm GeV}$, $\rm p_{T}^{j2} > 150~{\rm GeV}$, 
$\rm p_{T}^{j3} > 100~{\rm GeV}$, $\rm p_{T}^{j4} > 50~{\rm GeV}$,
since the signal has a significantly larger number of hard jets as compared to the background.
\item C7: Additionally, we demand at least 2 b-tagged jets, to suppress backgrounds 
like $\rm W$ ($\rm \to l\nu$)+jets. 
\item C8: A single isolated lepton is required with $\rm p_{T}^{\ell} \ge 20~{\rm GeV}$ and $|\rm \eta_{\ell}| \le 2.5$.
\item C9 : Since the leptonic mode is under investigation, a large $\PMET$ 
is required to suppress backgrounds like $\rm {t\bar{t}}$ + jets, $\rm {t\bar{t}H_{SM}}$. 
Therefore a $\PMET$ $> 400~{\rm GeV}$ is imposed. 
\item C10 : Due to the significantly larger jet activity in the signal, we impose 
$\rm H_{T}= \Sigma \rm p_{T}^{j} > 600~{\rm GeV}$, where jets  
satisfying the basic selection criteria are considered here.  
\item C11 : The SM process  $\rm t\bar{t}H_{SM}$, with one of the tops decaying 
leptonically is the largest background. To suppress this, we demand 
that the transverse mass of the lepton and missing transverse energy of the system, 
$\rm M_{T}(\ell, \PMET)= \sqrt{2 p_{T}^{\ell}\times \PMET\times(1- \cos\phi(\ell,\PMET) )} > 200~{\rm GeV}$.  
This ensures that the single lepton final states originating from the $\rm W$ boson 
are suppressed as the end point of the $\rm M_{T}(\ell,\PMET)$ distribution is 
expected to be bounded from above by the mass of the $\rm W$ boson. However, for 
the signal the presence of the additional $\chi_{1}^{0}$ ensures 
that the end point of the $\rm M_{T}(\ell, \PMET)$ distribution is shifted 
beyond the $\rm W$ boson mass.
\item C12: Events with number of jets greater than six are accepted. 
\end{itemize}

\item {\bf{Case II (Di-lepton)}}
The di-lepton channel, although being cleaner compared to the single lepton channel, 
has the disadvantage of a lower signal cross section. We optimize the 
di-lepton channel by imposing the following set of cuts.

\begin{itemize}
\item C13 : We demand at least 3 jets with 
$\rm p_{T}^{j1} > 150~{\rm GeV}$, $\rm p_{T}^{j2} > 100~{\rm GeV}$, $\rm p_{T}^{j3} > 50~{\rm GeV}$.
\item C14 : At least two b-tagged jets are required. 
\item  C15 : We demand two isolated leptons with $\rm p_{T}^{\ell} > 20~{\rm GeV}$ and $|\rm \eta_{\ell}| < 2.5$.
\item C16 : In order to suppress the backgrounds like $\rm t\bar{t}$ + jets, 
$\rm t\bar{t}H_{SM}$, which yield di-leptons from the $\rm W$ decay, we 
use the variable $\rm M_{T2}$\cite{Lester:1999tx,Barr:2003rg}, 
 defined as, 
{
\begin{equation} 
\rm M_{T2}(\PTV^{\ell1}, \PTV^{\ell2},{\PMETV}) \; = \; \rm 
min_{_{_{_{\hspace{-1cm}\Large {\PMETV} = 
\rm \PMETV^{\, 1} + \rm \PMETV^{\, 2}}}}} 
\bigg[max\{M_T(\PTV^{\ell1},{\PMETV^{\,1}}), 
M_T(\PTV^{\ell2},{\PMETV^{\, 2}})\}\bigg],
\label{mt2def} 
\end{equation} 
}
where 
$\ell1$ and $\ell2$ are the two isolated leptons, 
$\rm \PMET$ is the total missing transverse momentum of 
the event and $\rm M_T(\PTV^{v_1}, \PTV^{v_2})$ is 
the transverse mass of the system, defined as $$\rm 
M_T(\PTV^{v_1}, \PTV^{v_2}) = \sqrt{2 
|\PTV^{v_1}|\,|\PTV^{v_2}| (1 - 
\cos\phi)},$$ $\phi$ being the (azimuthal) angle between 
$\PTV^{v_1}$ and $\PTV^{v_2}$.
In the definition of Eq.\ref{mt2def}, $\rm \PMETV^{\, 1}$ and $\rm \PMETV^{\, 
2}$ are the two hypothetical splits of the total missing transverse 
momentum. It is assumed that the mass of the invisible particles are zero \cite{Barr:2009wu}.
The use of di-leptonic $\rm M_{T2}$ to suppress SM backgrounds has been considered
previously in a host of other works, for example see 
Refs.~\cite{Plehn:2010st,Plehn:2012pr,Chakraborty:2013moa,Belanger:2013oka}. 
For leptons originating 
from the decay of the top quark, the value of $\rm M_{T2}$ is bounded by 
the $W$ mass, whereas for signal the presence of $\chi_{1}^{0}$ 
ensures that the distribution is shifted beyond the $W$ mass. The 
signal and background distribution illustrating this feature is presented 
in Fig. \ref{mt2}. In this work, we set $\rm M_{T2} > 150~{\rm GeV}$.

\item C17 : Additionally we impose a criteria on the total missing transverse momentum, 
$\PMET$ $> 400~{\rm GeV}$. 
\end{itemize}

\end{enumerate}

\section{Results}
\label{sec4}


In this section, we summarize our findings based on the collider 
simulation described in the previous section. 
We begin our discussion with the the jet substructure analysis.
The event summary for the signal after individual selection cuts are 
presented in Table \ref{tab3}, while in Table \ref{tab4} the event summary for the 
backgrounds are tabulated.
The second column of Table \ref{tab3} 
corresponds to the NLO pair production cross-section of the 
light stop, as obtained from the official LHC supersymmetry 
cross section working group \cite{Borschensky:2014cia},
 corresponding to the benchmark points in Table \ref{tab2}. The
cross sections for $\rm t\bar{t}$ + jets are obtained using Madgraph5 
with an appropriate K-factor for 13 {\rm TeV} LHC, while 
for the cross section of $\rm t\bar{t} H_{SM}$ process we use LHC 
Higgs cross-section working group report \cite{lhchiggs}. 
Although all the other backgrounds like $\rm W/Z$+jets, $\rm t\bar{t}b\bar{b}$,
$\rm t\bar{t}Z$, $\rm t\bar{t}W$ as mentioned in the previous section were
simulated, we present only $\rm t\bar{t}$ + jets and  $\rm t\bar{t} H_{SM}$, 
as these are the dominant ones. It was checked that the choices 
of the selection cuts are such that other processes do 
not contribute to the final background cross section.

\begin{table}[!ht]
\small
\begin{center}
\tabulinesep=1.2mm
\begin{tabu}{|c|c|c|c|c|c|c|c|} 
\hline 
\multicolumn{2}{|}{} & 
\multicolumn{5}{|c|}{Effective cross-section after the cuts (in fb)}  \\
\hline
Signal & Production               
& C1 & C2   & C3 & C4 &  C5  \\
       &  c.s. (fb)  & &    
        &      &    &        \\
\hline 
P1 &  27.2   & 1.18 & 0.58 & 0.48  & 0.25 & 0.146  \\ 
P2 &  12    & 0.42 & 0.21 & 0.17 & 0.11 & 0.0794   \\
P3 &   6    &  0.22 & 0.11 & 0.09 & 0.07 & 0.0529  \\
P4 &   1.5   & 0.0497 & 0.0262 & 0.0225 & 0.0193 & 0.0161  \\
P5 &    0.5  &  0.0258 & 0.0139 & 0.0121 & 0.0115 & 0.0096  \\
\hline 
\end{tabu}
\caption{{\bf Jet Substructure}: Event summary for the signal after 
individual cuts as described in the text. The simulation is performed 
at 13 {\rm TeV} LHC energy. From 3rd column, we show 
the effective cross-section (c.s.) after all the event selection cuts have been 
applied.}
\label{tab3} 
\end{center}
\end{table} 
\begin{table}[!ht]
\small
\begin{center}
\tabulinesep=1.2mm
\begin{tabu}{|c|c|c|c|c|c|c|c|}
\hline
\multicolumn{2}{|}{} &
\multicolumn{5}{|c|}{Effective cross-section after the cuts (in fb)}  \\
\hline
SM bkgs & Production
& C1 & C2   & C3 & C4 &  C5  \\
       &  c.s. (fb)  & &
        &      &    &        \\
\hline
$\rm t \bar{t} + $ jets &
 700000 & 751.02  & 554.85  & 7.5  & 0  & 0 \\
$\rm t t H_{SM}$ &
500 & 14.45 & 9.76 & 7.47  & 0.5  & 0.028  \\
\hline 
Total & \multicolumn{5}{|c|}{} &  \\
bkg & \multicolumn{5}{|c|}{} & 0.028\\
\hline
\end{tabu}
\caption{{\bf Jet Substructure}: Event summary for the backgrounds (bkgs) after 
individual cuts as described in the text. Numbers in the last column show 
the final cross-section (c.s.) after all the event selection cuts are 
applied. For the $\rm t \bar{t} + $ jets background, we generate 
a matched sample of $\rm t \bar{t} + $ 0 jet, $\rm t \bar{t} + $ 1 jet and 
$\rm t \bar{t} + $ 2 jets using Madgraph. We also consider $W$+jets and $Z$+jets 
backgrounds, however both of them are identically zero at the end of C5.}
\label{tab4} 
\end{center}
\end{table} 


 \begin{table}[ht!]
 \small
 \begin{center}
 \tabulinesep=1.2mm
 \begin{tabu}{|c|c|c|c|c|c|c|c|} 
 \cline{3-8} 
 \multicolumn{1}{c}{} & 
 \multicolumn{1}{c|}{}& 
  \multicolumn{3}{|c|}{Signal($\rm N_S$) ( Background($\rm N_B$))} & 
  \multicolumn{3}{|c|}{{$ \rm{Significance}(\mathcal S) \; \rm{for} \;  \kappa = 10\% $}} \\ 
 \hline
   & $ \rm m_{\tilde{t}_1}$({\rm GeV}) &  100 $\rm fb^{-1}$ & 300 $\rm fb^{-1}$ & 1000 $\rm fb^{-1}$ & 100 $\rm fb^{-1}$ 
& 300 $\rm fb^{-1}$ & 1000 $\rm fb^{-1}$ \\ 
 \cline{1-8}
  P1 & 804.3 &14.6 (2.8) &43.8 (8.4)&146 (28) & 8.6 & 14.5 & 24.37 \\ 
  P2 & 908.2 &7.94 (2.8) &23.82 (8.4)&79.4 (28) &2.75 & 7.9  & 13.25  \\ 
  P3 & 1003.7 &5.29 (2.8) &15.87 (8.4)&52.9 (28) &1.83 & 5.25  & 8.83  \\ 
  P4 & 1211.5 &1.61 (2.8) & 4.83 (8.4)&16.1 (28) &0.55 & 1.6  & 2.69  \\ 
  P5 & 1392.5 &0.96 (2.8) & 2.88 (8.4) &9.6 (28) &0.33 & 0.95  & 1.6 \\ 
  \hline 
 \end{tabu}
  \caption{{\bf Jet Substructure:} The summary of our signal and backgrounds. Columns 3-5 show the number of signal 
  (total background) events for three values of the integrated luminosity: 100 $\rm fb^{-1}$, 300 $\rm fb^{-1}$ and 
  1000 $\rm fb^{-1}$ at 13 {\rm TeV} LHC.
  The columns 6-8 show the statistical significance of our signal for the above three 
  integrated luminosities. For each value of the integrated luminosity the significances are shown considering 
  $\kappa=$ 10\% systematic uncertainty.} 
 \label{tab5} 
  \end{center}
 \end{table} 

 From the 3rd column to the 7th column (C1-C5), 
we present the cut flow, normalized to the cross sections after  
imposing the selection cuts, as discussed in the previous section.  
Similarly, in Table \ref{tab4}, we discuss the event summary for the 
backgrounds after each cut. {As can be observed 
from the 3rd column of Table \ref{tab3}, about 3-6\% percent 
of the events have at least one tagged Higgs 
for the signal events.
The $\rm t\bar{t}$ + jets background is reduced by 99.89\% 
(see Table \ref{tab4}, 3rd column) while 
backgrounds like W/Z+jets, $\rm t\bar{t}b\bar{b}$,
$\rm t\bar{t}Z$, $\rm t\bar{t}W$ (not shown in the Table
) are reduced to negligible amounts. The largest background, 
as expected is $\rm t\bar{t}H_{SM}$, with a tagging efficiency 
of about 2.8\%. Therefore to reduce this background  the subsequent 
cuts C2-C5 play an important role. From Table \ref{tab4}, we 
notice that the principal cut to suppress the rest of the $\rm t\bar{t}$ +jets 
background is the demand of extra b-jets (cut C3) after satisfying 
the criteria of at least one SM-like Higgs. As mentioned earlier, 
these extra b-jets are unlikely to be present in the background and we 
observe that about 98\% of the remaining  $\rm t\bar{t}$ +jets background
is further rejected. In signal and $\rm t\bar{t}H_{SM}$ however these extra b-jets
are naturally present, and therefore they are not significantly suppressed.
To suppress the $\rm t\bar{t}H_{SM}$ background, we employ the cut on $\rm H_{T}$ and $\rm \PMET$
as noted in the previous section. Even after the Higgs tag the presence of 
significantly high energetic jets from the stop decay as compared to $\rm t\bar{t}H_{SM}$ 
and $\rm t\bar{t}$ +jets ensures that the $\rm H_{T}$ distribution 
is extended well beyond 1 {\rm TeV}. The optimized $\rm H_{T}$ cut of 800 {\rm GeV} 
kills the $\rm t\bar{t}H_{SM}$ background by about 93\%, while 
the $\rm t\bar{t}$ +jets background is entirely wiped out. After all cuts, the 
only surviving background turns out to be $\rm t\bar{t}H_{SM}$ as expected, with 
the background cross section being 0.028 fb.

The cut efficiency after all the selection cuts, in the jet 
substructure scenario, increases steadily from 
the benchmark points P1 to P5 due to a slight increase in the Higgs 
tagging efficiency as well as an increase in the strength of 
$\rm H_{T}$ and $\PMET$ cuts. However the rapid fall of 
the stop pair production cross-section with increasing mass 
results in a minuscule final cross-section for stop masses 
above 1 {\rm TeV}. To project the reach at 13 {\rm TeV}, we define the 
signal significance ($\mathcal S$) as 
$\rm S/\sqrt{(B+ (\kappa B)^2)}$, where we include a systematic uncertainty 
factor of $\rm \kappa$ in the background estimation to compensate 
for the fact that we do not perform any dedicated detector 
simulation in this study.
For this study we choose $\kappa$ to be 10$\%$. In Table \ref{tab5}, the 
signal significances for 100, 300, 1000 ${\rm fb}^{-1}$ of integrated luminosity are 
tabulated in columns 6-8 for the five benchmark points. It can be 
noted that one can probe a light stop mass up to 800~{\rm GeV}(P1) at 100 $\rm fb^{-1}$
integrated luminosity at 13 {\rm TeV} LHC with a significance $>$ 5, 
while stop masses up to 1 {\rm TeV} (P3) can be observed at 300 $\rm fb^{-1}$ 
luminosity. The signal suffers because of low production cross section, and 
only with the very high luminosity run at LHC (i.e 3000 $\rm fb^{-1}$),
 one can achieve signal significances greater than 5$\sigma$ for 
the benchmark points P4 and P5 with stop masses greater than 1 {\rm TeV}. 
We also observe that one could exclude $ \rm m_{\tilde{t}_1}$ $\sim$ 1.2 {\rm TeV} 
for 300 $\rm fb^{-1}$ integrated luminosity with the jet substructure technique.

\begin{table}[ht!]
\small
\begin{center}
\tabulinesep=1.2mm
\begin{tabu}{|c|c|c|c|c|c|c|c|c|} 
\hline 
\multicolumn{2}{|}{} & 
\multicolumn{7}{|c|}{Effective cross-section after the cuts (in fb)} \\
\hline
Signal & Production               
& C6 & C7   & C8 & C9 &  C10  & C11 & C12 \\
       &  c.s. (fb)  & &    
        &      &    &    &    &     \\
\hline 
P1 &  27.2  & 16.39 & 10.44 & 3.55  & 0.442 & 0.441  & 0.27 & 0.239 \\ 
P2 &  12  & 8.49 & 5.5 & 1.88 & 0.315 & 0.314 & 0.2 & 0.176   \\
P3 &   6  &  4.34 & 2.54 & 0.88 & 0.259 & 0.259 & 0.169 & 0.139  \\
P4 &   1.5  & 1.17 & 0.74 & 0.246 & 0.109 & 0.109 & 0.077 & 0.061  \\
P5 &    0.5 & 0.45 & 0.29 & 0.098 & 0.040 & 0.040 & 0.028  & 0.0255  \\
\hline 
\end{tabu}
\caption{{\bf One lepton}: Event summary for the signal after 
individual cuts as described in the text. In the last column we show 
the final cross-section (c.s.) after all the event selection cuts have been 
applied.}
\label{tab6} 
\end{center}
\end{table} 
\begin{table}[htb!]
\small
\begin{center}
\tabulinesep=1.2mm
\begin{tabu}{|c|c|c|c|c|c|c|c|c|}    
\hline
\multicolumn{2}{|}{} &
\multicolumn{7}{|c|}{Effective cross-section after the cuts (in fb)} \\
\hline
SM bkgs & Production                    
& C6 & C7   & C8 & C9 &  C10  & C11 & C12 \\
       &  c.s. (fb)  & &      
        &      &    &    &    &     \\
\hline 
$\rm t \bar{t} + $ jets &
 700000 & 20465.1  & 8130.2  & 1410.6  & 9.13  & 8.96   &  5.6 & 2.28 \\
$\rm t t H_{SM}$ &
500 & 54.75 & 37.12 & 8.1  & 0.089  & 0.089 & 0.054 & 0.036  \\
$\rm W + jets$ &
83315178 & 6153.2 & 0 & 0  & 0  & 0 & 0 & 0   \\
$\rm Z + jets$ &
42070505 & 4115.1 & 470.3 & 0  & 0  & 0 & 0 & 0   \\
\hline 
Total & \multicolumn{7}{|c|}{} &  \\
bkg & \multicolumn{7}{|c|}{} & 2.316\\
\hline 
\end{tabu}
\caption{{\bf One lepton}: Event summary for the backgrounds (bkgs) after 
individual cuts as described in the text. In the last column we show 
the final cross-section (c.s.) after all the event selection cuts have been 
applied. For the $\rm t \bar{t} + $ jets background we have generated 
a matched sample of $\rm t \bar{t} + $ 0 jet, $\rm t \bar{t} + $ 1 jet and 
$\rm t \bar{t} + $ 2 jets using Madgraph.}
\label{tab7} 
\end{center}
\end{table} 
 \begin{table}[ht!]
 \small
 \begin{center}
 \tabulinesep=1.2mm
 \begin{tabu}{|c|c|c|c|c|c|c|c|}
 \cline{3-8}
 \multicolumn{1}{c}{} &
 \multicolumn{1}{c|}{}&
  \multicolumn{3}{|c|}{Signal($\rm N_S$) ( Background($\rm N_B$))} &
  \multicolumn{3}{|c|}{{$ \rm{Significance}(\mathcal S) \; \rm{for} \;  \kappa = 10\% $}} \\
 \hline
   & $ \rm m_{\tilde{t}_1}$({\rm GeV}) &  100 $\rm fb^{-1}$ & 300 $\rm fb^{-1}$ & 1000 $\rm fb^{-1}$ & 100 $\rm fb^{-1}$ 
& 300 $\rm fb^{-1}$ & 1000 $\rm fb^{-1}$ \\ 
 \cline{1-8}
  P1 & 804.3 & 23.9 (231.6) & 71.7 (694.8)& 239 (2316) & 0.86 & 0.96 & 1.01  \\ 
  P2 & 908.2 & 17.6 (231.6) & 52.8 (694.8)& 176 (2316) &  0.64  & 0.71  & 0.74  \\ 
  P3 & 1003.7 & 13.9 (231.6) & 41.7 (694.8)& 139 (2316) & 0.51 &  0.56 & 0.58  \\ 
  P4 & 1211.5 & 6.1 (231.6) & 18.3 (694.8)& 61 (2316) &  0.22 &  0.25 & 0.26  \\ 
  P5 & 1392.5 & 2.55 (231.6) & 7.65 (694.8) & 25.5 (2316) & 0.09 & 0.10  & 0.11  \\ 
  \hline
 \end{tabu}
  \caption{{\bf One lepton}: The summary of our signal and backgrounds. Columns 3-5 show the number of signal 
  (total background) events for three values of the integrated luminosity: 100 $\rm fb^{-1}$, 300 $\rm fb^{-1}$ and 
  1000 $\rm fb^{-1}$. The columns 6-8 show the statistical significance of our signal for the above three 
  integrated luminosities. For each value of the integrated luminosity the significance is shown for three choices 
  of the amount of possible systematic uncertainties, $\kappa=$ 10\%. }
 \label{tab8}
  \end{center}
 \end{table}

We compare the results obtained using the jet substructure method with the leptonic searches for this 
 channel. We first analyze the case of single lepton as denoted by Case-I in 
the previous section. In Table \ref{tab6} and Table \ref{tab7}, we summarize 
the event yields (normalized to cross section) for the signal and backgrounds respectively.
 We find that the $\rm W$+ jets background 
is entirely killed by the demand of b-tagged jets, as expected. For the $\rm Z$ + jets background, 
the demand of two b-tagged jets, along with four hard jets implies that 
no isolated lepton is expected to be present. Therefore, imposition of the single lepton criteria (see  column C8, Table \ref{tab7}) eliminates the Z+ jets background. The remaining background consists of 
$\rm t\bar{t}$ + jets and $\rm t\bar{t}H_{SM}$. As can be seen 
from column 9 in Table \ref{tab7}, $\rm \PMET$ along with $\rm H_{T}$ and $\rm M_{T}$ suppresses 
this set to a negligible level. However the small starting 
cross section for the signal essentially leads to the fact that the 
signal significance is not large enough for a discovery (or an 
exclusion) even at the 13 {\rm TeV} high luminosity scenario for these benchmark 
points as can be observed from Table \ref{tab8}.

We next turn our attention to the di-lepton channel, denoted 
by case-II in the previous section. The signal and background yields 
in this case are tabulated in Table \ref{tab9} and 
Table \ref{tab10} respectively corresponding to the selection 
cuts C13 - C17 described in the last section. 
The demand for two isolated leptons in cut C13 suppresses the 
$W$ + jets background, while rest of the background is suppressed 
by the $\rm M_{T2}$ cut (C16). The $\rm M_{T2}$ cut is expected to 
be bounded from above by the mass of the $W$ boson for the background, while for signal
it is expected to extend to higher values due to the presence of 
large missing energy. We observe that the backgrounds 
like $\rm t\bar{t}$ + jets is reduced to zero (0) after the $\rm M_{T2}$
cut, while there is a small residue from the $\rm t\bar{t}H_{SM}$
background, which are further reduced by the $\PMET$ cut of C17. 
In Table \ref{tab11}, we observe that for the di-lepton channel, 
up to 1 {\rm TeV} stops can be discovered with 100 $\rm fb^{-1}$ integrated 
luminosity, while a light stop mass up to 1.2 {\rm TeV} can be discovered with 
300 $\rm fb^{-1}$ of luminosity. Therefore we conclude that the 
di-lepton channel is most suited to probe the heavier $\rm \tilde{t}_{1}$ 
mass with the benchmark points depicted in Table \ref{tab2}.
We also observe that stop masses up to 1.4 {\rm TeV} can be excluded at 
13 {\rm TeV} LHC, with 300 $\rm fb^{-1}$ luminosity.

\begin{table}[htb!]
\small
\begin{center}
\tabulinesep=1.2mm
\begin{tabu}{|c|c|c|c|c|c|c|c|}
\hline
\multicolumn{2}{|}{} &
\multicolumn{5}{|c|}{Effective cross-section after the cuts (in fb)}  \\
\hline
Signal & Production
& C13 & C14   & C15 & C16 &  C17  \\
       &  c.s. (fb)  & &
        &      &    &        \\
\hline
P1 &  27.2  & 24.61 & 15.29 & 1.49  & 0.17 & 0.038 \\ 
P2 &  12  & 11.31 & 7.17 & 0.63 & 0.096 & 0.0264  \\
P3 &   6  &  5.65 & 3.19 & 0.297 & 0.061 & 0.0262  \\
P4 &   1.5 & 1.43 & 0.875 & 0.065 & 0.019 & 0.012  \\
P5 &    0.5 & 0.494 & 0.314 & 0.026 & 0.0069 & 0.0035  \\
\hline
\end{tabu}
\caption{{\bf Two lepton}: Event summary for the signal after 
individual cuts as described in the text. In the last column we show 
the final cross-section (c.s.) after all the event selection cuts have been 
applied.}
\label{tab9} 
\end{center}
\end{table} 

\begin{table}[htb!]
\small
\begin{center}
\tabulinesep=1.2mm
\begin{tabu}{|c|c|c|c|c|c|c|c|}
\hline
\multicolumn{2}{|}{} &
\multicolumn{5}{|c|}{Effective cross-section after the cuts (in fb)}  \\
\hline
SM bkgs & Production
& C13 & C14   & C15 & C16 &  C17  \\
       &  c.s. (fb)  & &
        &      &    &        \\
\hline
$\rm t \bar{t} + $ jets &
 700000 & 105014.18  & 38630.79  & 515.89  & 0  & 0  \\
$\rm t t H_{SM}$ &
500 & 179.56 & 116.06 & 3.01  & 0.007 & 0.001  \\
$\rm W + jets$ &
83315178 & 123064.92  & 0 & 0  & 0 & 0   \\
$\rm Z + jets$ &
42070505 & 78891.62 & 4799 & 0  & 0  & 0  \\
\hline 
Total & \multicolumn{5}{|c|}{} &  \\
bkg & \multicolumn{5}{|c|}{} & 0.001\\
\hline
\end{tabu}
\caption{{\bf Two lepton}: Event summary for the backgrounds (bkgs) after 
individual cuts as described in the text. In the last column we show 
the final cross-section (c.s.) after all the event selection cuts have been 
applied. For the $\rm t \bar{t} + $ jets background we have generated 
a matched sample of $\rm t \bar{t} + $ 0 jet, $\rm t \bar{t} + $ 1 jet and 
$\rm t \bar{t} + $ 2 jets using Madgraph.}
\label{tab10} 
\end{center}
\end{table} 

 \begin{table}[ht!]
 \small
 \begin{center}
 \tabulinesep=1.2mm
 \begin{tabu}{|c|c|c|c|c|c|c|c|}
 \cline{3-8}
 \multicolumn{1}{c}{} &
 \multicolumn{1}{c|}{}&
  \multicolumn{3}{|c|}{Signal($\rm N_S$) ( Background($\rm N_B$))} &
  \multicolumn{3}{|c|}{{$ \rm{Significance}(\mathcal S) \; \rm{for} \;  \kappa = 10\% $}} \\
 \hline
   & $ \rm m_{\tilde{t}_1}$({\rm GeV}) &  100 $\rm fb^{-1}$ & 300 $\rm fb^{-1}$ & 1000 $\rm fb^{-1}$ & 100 $\rm fb^{-1}$
& 300 $\rm fb^{-1}$ & 1000 $\rm fb^{-1}$ \\
 \cline{1-8}
  P1 & 804.3 & 3.8 (0.1) & 11.4 (0.3) & 38 (1) & 12.1 & 20.8  & 37.8 \\
  P2 & 908.2 & 2.64 (0.1) & 7.92 (0.3)& 26.4 (1) & 8.34 & 14.4 & 26.3  \\
  P3 & 1003.7 & 2.62 (0.1)  & 7.86 (0.3)& 26.2 (1) & 8.28 & 14.3  &26.0    \\
  P4 & 1211.5 & 1.2 (0.1) & 3.6 (0.3)& 12.0 (1) & 3.79 & 6.56  & 11.9  \\
  P5 & 1392.5 & 0.35 (0.1) & 1.1 (0.3) & 3.5 (1) & 1.1 & 2.0  & 3.48 \\
  \hline
 \end{tabu}
  \caption{{\bf Two lepton}: The summary of our signal and backgrounds. Columns 3-5 show the number of signal
  (total background) events for three values of the integrated luminosity: 100 $\rm fb^{-1}$, 300 $\rm fb^{-1}$ and
  1000 $\rm fb^{-1}$. The columns 6-8 show the statistical significance of our signal for the above three
  integrated luminosities. For each value of the integrated luminosity the significance is shown for three choices
  of the amount of possible systematic uncertainties, $\kappa=$ 10\%. }
 \label{tab11}
  \end{center}
 \end{table}

Finally, we examine the possibility of the detection of
multiple Higgs peaks in the fat-jet mass ($\rm m_{J}$) distribution over
the SM background, where $\rm J$ denotes the fat-jet. In Fig. \ref{minv},
we plot the fat-jet distribution for the jet substructure
technique at 300, 1000, and 3000 $\rm fb^{-1}$ luminosity respectively. 
We plot first six $\rm p_T$-ordered fat-jet masses at the end of 
our selection cut (C5) in our jet substructure Higgs analysis. In these plots 
we demand two b-tagged jets only for the fat-jet satisfying 
the 125 {\rm GeV} Higgs criteria, i.e, 
$\rm 110 < m_{{J}} < 140 ~ {\rm GeV}$. 
We observe that while the SM Higgs peak is overwhelmingly visible at
all the luminosity options, one can also observe 
slight excesses in three different masses over 
the $\rm t\bar t H_{SM}$ background. We find small excesses in the 
lighter CP even Higgs boson  ($\sim$ 65 {\rm GeV}), $Z$ boson 
 ($\sim$ 91 {\rm GeV}) as well as at the top quark 
($\sim$ 172 {\rm GeV}) masses. With a better branching ratio 
of $ \chi_{i}^{0}\to \chi_{1}^{0}H_{1}$ ($i = 2,..,5$), it
is possible that the peak corresponding to the lighter CP even Higgs 
might be visible at even lower luminosities.} However, 
for the benchmark points of our choice which are not the most ideal
for this purpose as the branching ratio $\rm \chi_{2}^{0}\to \chi_{1}^{0}H_{1}$ is not
large enough, this possibility can only be realized at very high luminosity
options. Using a simple number counting method we observe that
with 3000 $\rm fb^{-1}$ luminosity one can achieve a 3$\sigma$ signal
significance to observe the light CP-even Higgs boson ($H_1$) for our 
representative benchmark point P3 with stop mass around 1 {\rm TeV}. The possibility of
having significant BR for the decay of $\rm \chi^{0}_{2,3}$ to
both the Higgses ($\rm H_{1,SM}$) is beyond the scope of this paper,
and we try to report this possibility elsewhere \cite{inprep}.

\begin{figure}[!htb]
\begin{center}
\includegraphics[angle =0, width=0.32\textwidth]{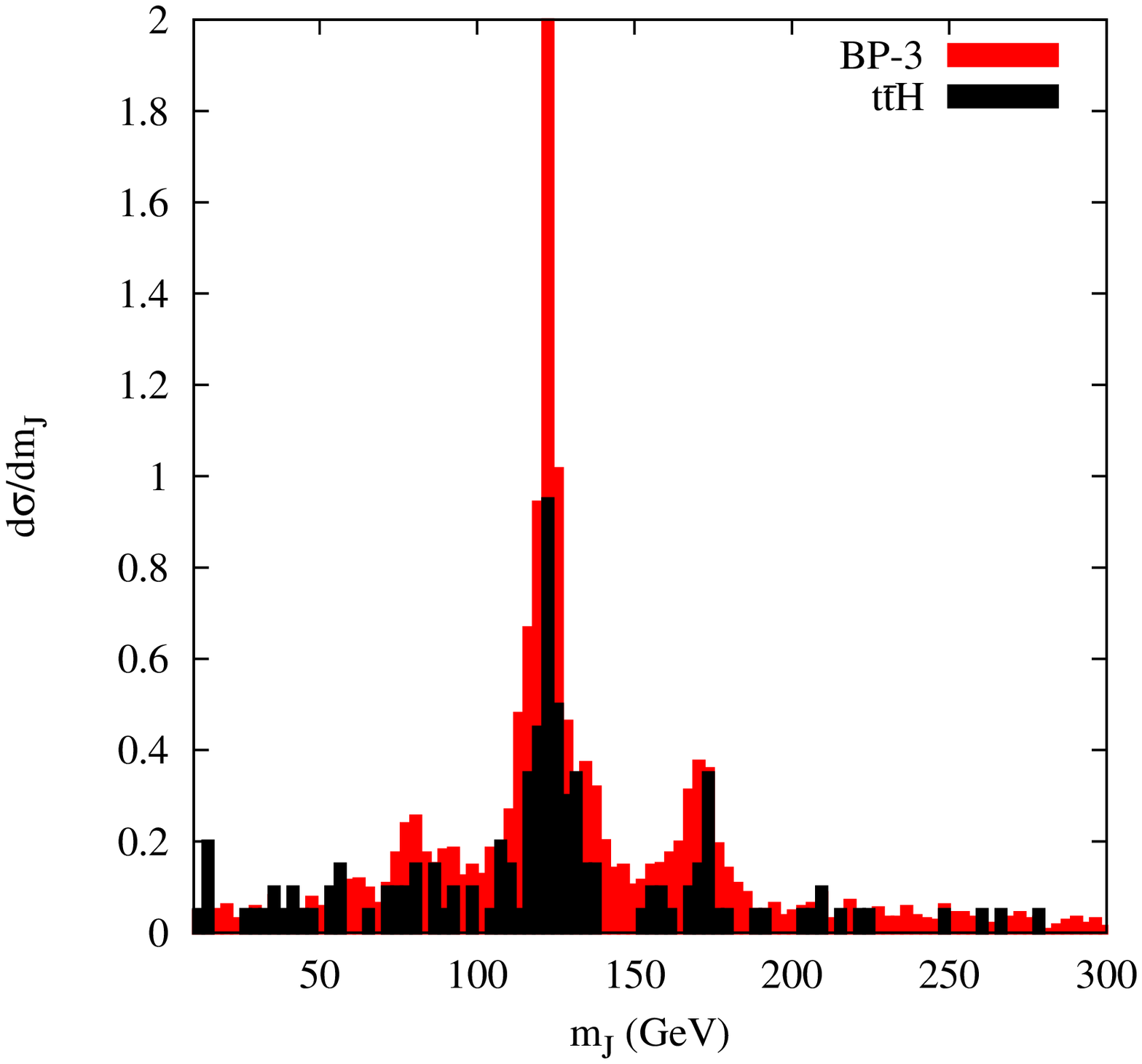}
\includegraphics[angle =0, width=0.32\textwidth]{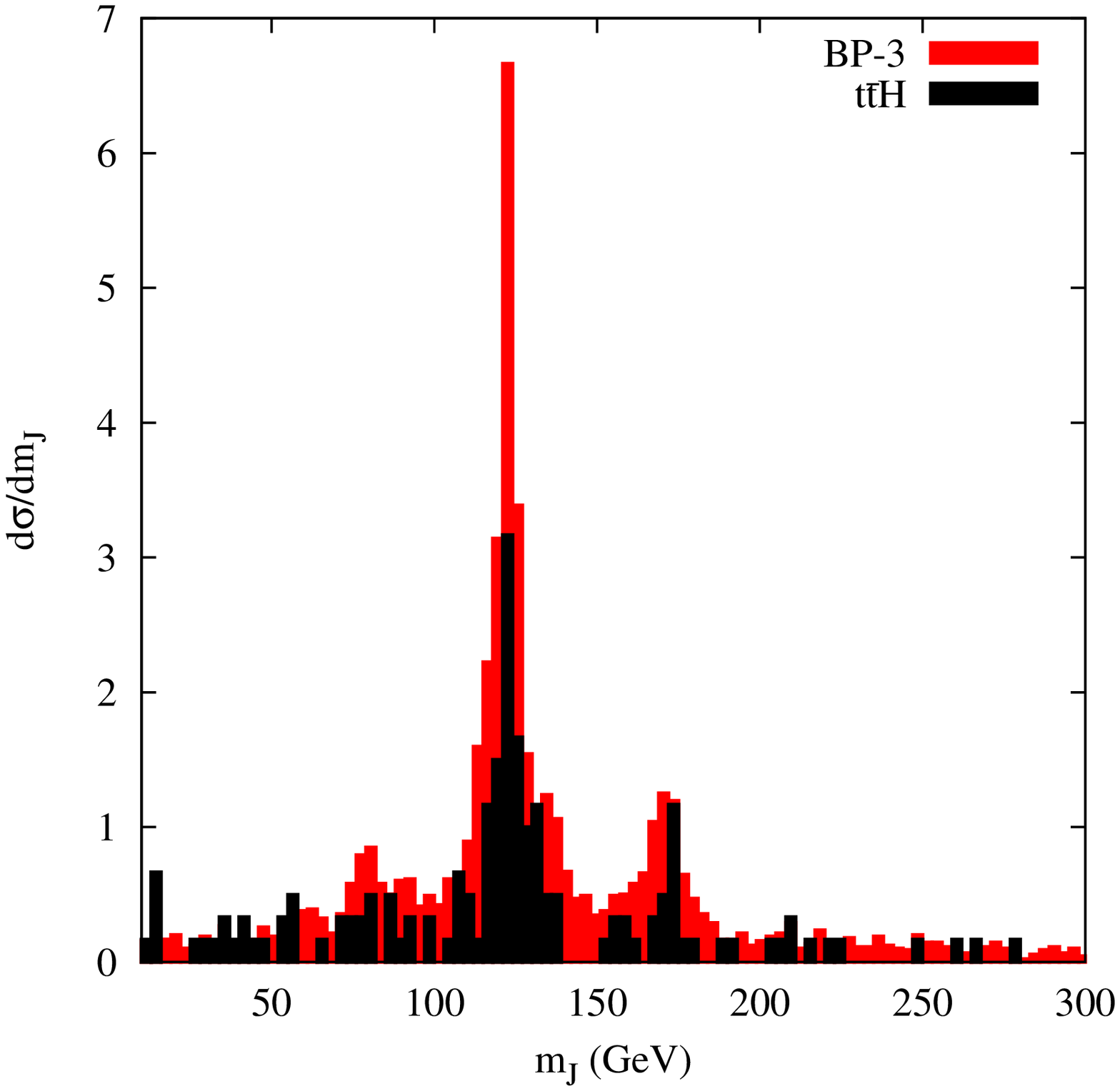}
\includegraphics[angle =0, width=0.32\textwidth]{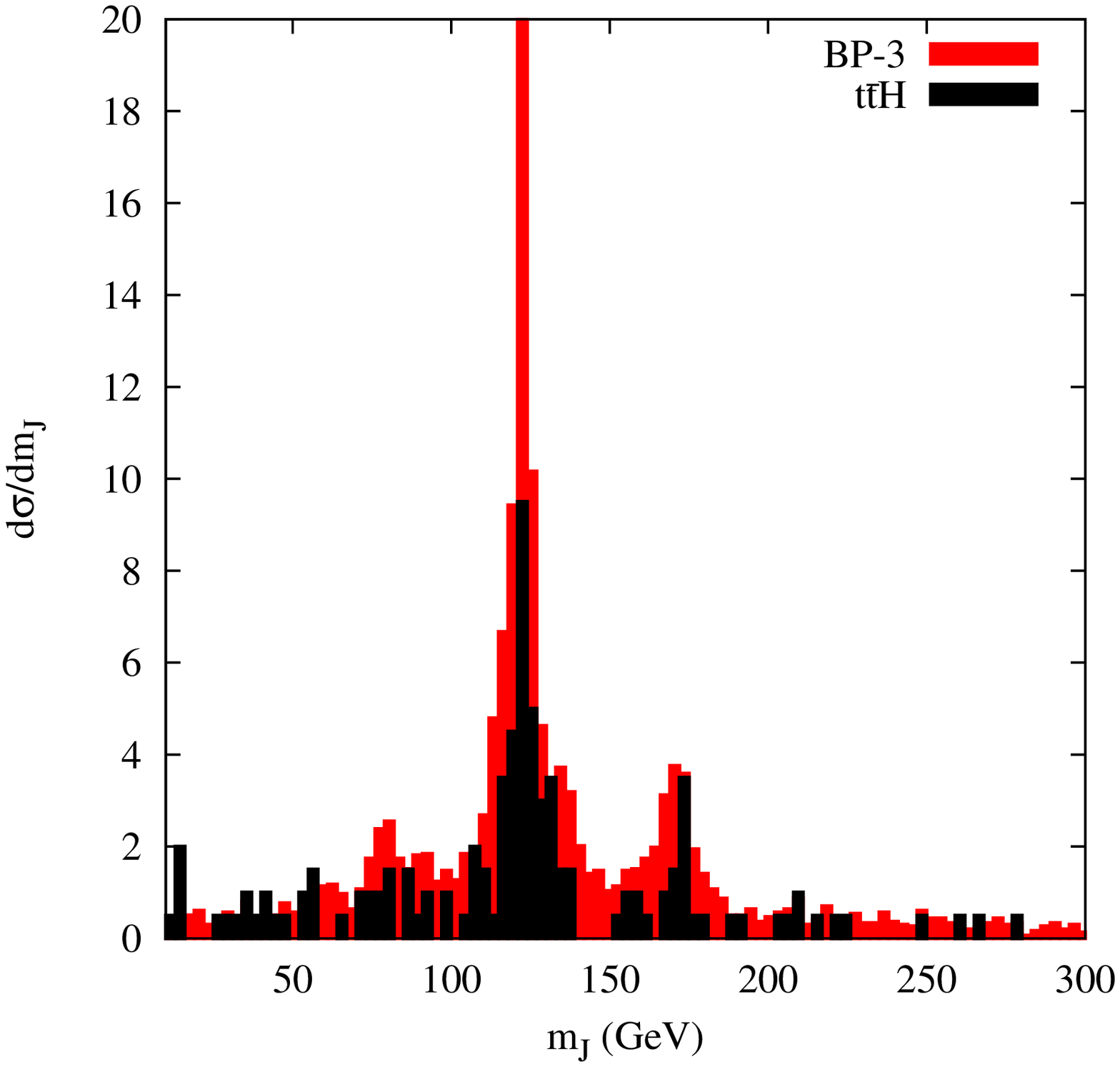}
\vspace*{-1.5cm}
\caption{ { Fat-jet mass distribution with 300 (left), 1000 (middle), 3000 
(right) $\rm fb^{-1}$ luminosity,
after all cuts as described by the jet substructure analysis
in the text. Here we plot first six $\rm p_T$-ordered fat-jet masses at 
the end of our selection cut (C5), no extra b-tagging criteria 
on these fat-jets are imposed expect the SM Higgs like fat-jets 
where we demand two b-tagged jets inside the fat-jet. The reason 
behind this plot is to investigate the possibility of having multiple 
Higgs peaks in the fat-jet mass distributions. 
The figures are all normalized to cross section with a particular 
choice of luminosity.
}}
\label{minv}
\end{center}
\end{figure}


\section{Summary}
\label{sec5}

In this article, we studied  Higgs signatures 
from the cascade decay of the light stop ($\widetilde{t_1}$) 
in the framework of NMSSM. We 
identify regions of parameter space where this channel 
is dominant and consistent with current Higgs
coupling measurements, LHC constraints, dark matter relic density and
direct detection constraints, and provide a 
detailed collider simulation of the signal and background. 
We probe this channel  using the 
jet substructure techniques and via the conventional leptonic searches. 
It is concluded that in terms of 
signal to background ratio, the di-lepton channel is the most 
promising one. In the di-lepton channel we can discover stop masses 
up to 1.2 {\rm TeV} with 300 $\rm fb^{-1}$ luminosity.  With the jet substructure method, 
stop masses up to 1 {\rm TeV} could be probed with 300 $\rm fb^{-1}$ luminosity. 
Finally we also investigate the possibility of the appearance of multiple 
Higgs peaks in the fat-jet mass distribution. This is plagued by the 
low branching ratio to the lighter CP even state in our case, and is 
only visible at very high luminosity run of the LHC.

\section{Acknowledgements}
AC would like to thank the Department of Atomic Energy, 
Government of India for financial support. The 
work of SM was partially supported by funding available from the 
Department of Atomic Energy, Government of India, for the Regional 
Centre for Accelerator-based Particle Physics
(RECAPP), Harish-Chandra Research Institute. DS acknowledges 
Sabine Kraml for a careful reading and valuable suggestions for the manuscript. DS and AC 
also acknowledges the help of Benjamin Fuks for his help with issues 
related to NMSSMTools.
 The work of DS is supported 
by the French ANR project DM-AstroLHC. DS also acknowledges the hosiptality 
of the Indian Associaltion for the Cultivation of Sciences, Kolkata, 
where part of the work was completed.


\bibliography{NMSSM-stop.bib}{}
\bibliographystyle{utphys.bst}

\end{document}